


\catcode`\@=11

\chardef\f@ur=4
\chardef\l@tter=11
\chardef\@ther=12
\dimendef\dimen@iv=4
\toksdef\toks@i=1 
\toksdef\toks@ii=2
\newtoks\emptyt@ks 

\def\glet{\global\let}
\def\gz@#1{\global#1\z@}
\def\gm@ne#1{\global#1\m@ne}
\def\g@ne#1{\global\advance#1\@ne}

\def\@height{height}       
\def\@depth{depth}         
\def\@width{width}         

\def\@plus{plus}           
\def\@minus{minus}         


\def\loop#1\repeat{\def\iter@te{#1\expandafter\iter@te \fi}\iter@te
  \let\iter@te\undefined}

\def\bpargroup{\bp@rgroup\ep@r} 
\def\bgrafgroup{\bp@rgroup\ep@rgroup} 
\def\bp@rgroup{\bgroup \let\par\ep@rgroup \let\endgraf}

\def\ep@r{\ifhmode \unpenalty\unskip \fi \p@r}
\def\ep@rgroup{\ep@r \egroup}

\let\p@r=\endgraf   
\let\par=\ep@r
\let\endgraf=\ep@r

\def\lb{\hfil\break}
\def\endpage{\par \vfil \eject}
\def\superendpage{\par \vfil \supereject}



\def\leftline{\@line\hsize\empty\hss}
\def\rightline{\@line\hsize\hss\empty}
\def\centerline{\@line\hsize\hss\hss}

\let\plainrlap=\rlap   
\let\plainllap=\llap   

\def\rlap{\@line\z@\empty\hss}
\def\llap{\@line\z@\hss\empty}

\def\lftline{\@line\hsize\empty\hfil}
\def\rtline{\@line\hsize\hfil\empty}
\def\ctrline{\@line\hsize\hfil\hfil}

\def\@line#1#2#3{\hbox to#1\bgroup#2\let\n@xt#3%
  \afterassignment\@@line \setbox\z@\hbox}
\def\@@line{\aftergroup\@@@line}
\def\@@@line{\unhbox\z@ \n@xt\egroup}

\def\after@arg#1{\bgroup\aftergroup#1\afterassignment\after@@arg\@eat}
\def\after@@arg{\ifcat\bgroup\noexpand\n@xt\else \n@xt\egroup \fi}
\def\@eat{\let\n@xt= } 
\def\eat#1{}           
\def\@eat@#1{\@eat}    


\let\nl=\space

\def\ctrlines#1#2{\par \bpargroup
  \bgroup \parskip\z@skip \noindent \egroup
  \let\ctr@style#1\let\nl\ctr@lines \hfil \ctr@style{#2}\strut
  \interlinepenalty\@M \par}
\def\ctr@lines{\strut \lb \strut \hfil \ctr@style}


\def\begin@{\ifmmode \expandafter\mathpalette\expandafter\math@ \else
  \expandafter\make@ \fi}
\def\make@#1{\setbox\z@\hbox{#1}\fin@}
\def\math@#1#2{\setbox\z@\hbox{$\m@th#1{#2}$}\fin@}

\def\ph@nt{\let\fin@\finph@nt \begin@}
\let\makeph@nt=\undefined
\let\mathph@nt=\undefined

\newif\ift@ \newif\ifb@
\def\topsmash{\t@true\b@false\sm@sh}
\def\botsmash{\t@false\b@true\sm@sh}
\def\smash{\t@true\b@true\sm@sh}
\def\sm@sh{\let\fin@\finsm@sh \begin@}
\let\makesm@sh=\undefined
\let\mathsm@sh=\undefined
\def\finsm@sh{\ift@\ht\z@\z@\fi \ifb@\dp\z@\z@\fi \box\z@}


\newdimen\boxitsep   \boxitsep=5pt

\def\fboxit#1#2{\hbox{\vrule \@width#1\p@
    \vtop{\vbox{\hrule \@height#1\p@ \vskip\boxitsep
        \hbox{\hskip\boxitsep #2\hskip\boxitsep}}%
      \vskip\boxitsep \hrule \@height#1\p@}\vrule \@width#1\p@}}


\begingroup
  \catcode`\:=\active
  \lccode`\*=`\\ \lowercase{\gdef:{*}}   
  \catcode`\;=\active
  \lccode`\* `\% \lowercase{\gdef;{*}}   
  \catcode`\^^M=\active \glet^^M=\space  
\endgroup


\begingroup
  \catcode`\:=\active
  \outer\gdef\comment{\begingroup
    \catcode`\\\@ther \catcode`\%\@ther \catcode`\^^M\@ther
    \catcode`\{\@ther \catcode`\}\@ther \catcode`\#\@ther
    \wlog{* input between `:comment' and `:endcomment' ignored *}%
    \c@mment}
\endgroup
{\lccode`\:=`\\ \lccode`\;=`\^^M
  \lowercase{\gdef\c@mment#1;{\c@@mment:endcomment*}}}
\def\c@@mment#1#2*#3{\if #1#3%
    \ifx @#2@\def\n@xt{\endgroup\ignorespaces}\else
      \def\n@xt{\c@@mment#2*}\fi \else
    \def\n@xt{\c@mment#3}\fi \n@xt}

\message{date and time,}


\newcount\langu@ge

\begingroup \catcode`\"=\@ther \gdef\dq{"}
            \catcode`\"=\active
            \gdef"#1{{\accent\dq 7F #1}\penalty\@M \hskip\z@skip}
  \endgroup

\outer\def\english{\gm@ne\langu@ge
  \global\catcode`\"\@ther \glet\3\undefined}
\outer\def\german{\gz@\langu@ge
  \global\catcode`\"\active \glet\3\ss}

\def\case@language#1{\ifcase\expandafter\langu@ge #1\fi}
\def\case@abbr#1{{\let\nodot\n@dot\case@language{#1}.^^>}}
\def\n@dot{\expandafter\eat}
\let\nodot=\empty


\def\themonth{\xdef\themonth{\noexpand\case@language
  {\ifcase\month \or Januar\or Februar\or M\noexpand\"arz\or April\or
  Mai\or Juni\or Juli\or August\or September\or Oktober\or November\or
  Dezember\fi
  \noexpand\else
  \ifcase\month \or January\or February\or March\or April\or May\or
  June\or July\or August\or September\or October\or November\or
  December\fi}}\themonth}

\def\thedate{\case@language{\else\themonth\ }\number\day
  \case@language{.\ \themonth \else ,} \number\year}

\def\date{\number\day.\,\number\month.\,\number\year}


\def\PhysTeX{$\Phi\kern-.25em\raise.4ex\hbox{$\Upsilon$}\kern-.225em
  \Sigma$-\TeX}


\count255=\time \divide\count255 by 60 \edef\thetime{\the\count255 :}
\multiply\count255 by -60 \advance\count255 by\time
\edef\thetime{\thetime \ifnum10>\count255 0\fi \the\count255 }

\newskip\refbetweenskip   \newskip\chskiptamount
\newskip\chskiplamount   \newskip\secskipamount
\newskip\footnotebaselineskip   \newskip\interfootnoteskip

\newdimen\chapstretch   \chapstretch=2.5cm
\newcount\chappenalty   \chappenalty=-800
\newdimen\sectstretch   \sectstretch=2cm
\newcount\sectpenalty   \sectpenalty=-400

\def\chskipt{\chapbreak \vskip\chskiptamount}
\def\chskipl{\nobreak \vskip\chskiplamount}
\def\unchskip{\vskip-\chskiplamount}
\def\secskipt{\sectbreak \vskip\secskipamount}
\def\chapbreak{\par \vskip\z@\@plus\chapstretch \penalty\chappenalty
  \vskip\z@\@plus-\chapstretch}
\def\sectbreak{\par \vskip\z@\@plus\sectstretch \penalty\sectpenalty
  \vskip\z@\@plus-\sectstretch}



\begingroup \lccode`\*=`\r
  \lowercase{\def\n@xt#1*#2@{#1}
    \xdef\font@sel{\expandafter\n@xt\fontname\tenrm*@}}\endgroup

\message{loading \font@sel\space fonts,}

=\font@sel r12 
=\font@sel r9
=\font@sel r8
=\font@sel r6

=\font@sel mi12 \skewchar\twelvei='177 
=\font@sel mi9    \skewchar\ninei='177
=\font@sel mi8   \skewchar\eighti='177
=\font@sel mi6   \skewchar\sixi='177

=\font@sel sy10 scaled \magstep1
  \skewchar\twelvesy='60 
=\font@sel sy9    \skewchar\ninesy='60
=\font@sel sy8   \skewchar\eightsy='60
=\font@sel sy6   \skewchar\sixsy='60




=\font@sel bx12 
=\font@sel bx9
=\font@sel bx8
=\font@sel bx6

=\font@sel tt12 
=\font@sel tt8


=\font@sel sl12 
=\font@sel sl9
=\font@sel sl8

\font\twelveit=\font@sel ti12 
\font\nineit=\font@sel ti9
\font\eightit=\font@sel ti8
=\font@sel ti7



=\font@sel csc10 scaled \magstep1 
=\font@sel csc10






\def\twelvepoint{\twelve@point
  \let\normal@spacing\twelve@spacing \set@spacing}
\def\tenpoint{\ten@point
  \let\normal@spacing\ten@spacing \set@spacing}
\def\eightpoint{\eight@point
  \let\normal@spacing\eight@spacing \set@spacing}

\def\rm{\fam\z@ \@fam}
\def\mit{\fam\@ne}
\def\oldstyle{\mit \@fam}
\def\cal{\fam\tw@}
\def\it{\fam\itfam \@fam}
\def\sl{\fam\slfam \@fam}
\def\bf{\fam\bffam \@fam}
\def\tt{\fam\ttfam \@fam}
\def\caps{\@caps}
\def\@fam{\the\textfont\fam}


\def\twelve@point{\set@fonts twelve ten eight }
\def\ten@point{\set@fonts ten eight six }
\def\eight@point{\set@fonts eight six five }

\def\set@fonts#1 #2 #3 {\textfont\ttfam\csname#1tt\endcsname
    \expandafter\let\expandafter\@caps\csname#1csc\endcsname
  \def\n@xt##1##2{\textfont##1\csname#1##2\endcsname
    \scriptfont##1\csname#2##2\endcsname
    \scriptscriptfont##1\csname#3##2\endcsname}%
  \set@@fonts}
\def\set@@fonts{\n@xt0{rm}\n@xt1i\n@xt2{sy}%
  \textfont3\tenex \scriptfont3\tenex \scriptscriptfont3\tenex
  \n@xt\itfam{it}\n@xt\slfam{sl}\n@xt\bffam{bf}\rm}


\def\singlespace{\chardef\@spacing\z@ \set@spacing}
\def\doublespace{\chardef\@spacing\@ne \set@spacing}
\def\triplespace{\chardef\@spacing\tw@ \set@spacing}
\chardef\@spacing=1   

\def\set@spacing{\expandafter\expandafter\expandafter\set@@spacing
  \expandafter\spacing@names\expandafter\@@\normal@spacing
  \normalbaselines}
\def\set@@spacing#1#2\@@#3+#4*{#1#4\multiply#1\@spacing \advance#1#3%
  \ifx @#2@\let\n@xt\empty \else
    \def\n@xt{\set@@spacing#2\@@}\fi \n@xt}

\def\normalbaselines{\lineskip\normallineskip
  \setbaselineskip\normalbaselineskip
  \lineskiplimit\normallineskiplimit}

\def\setbaselineskip{\afterassignment\set@strut \baselineskip}
\def\set@strut{\setbox\strutbox\spacer\z@\baselineskip}

\def\spacer{\hbox\bgroup \afterassignment\x@spacer \dimen@}
\def\x@spacer{\ifdim\dimen@=\z@\else \hskip\dimen@ \fi
  \afterassignment\y@spacer \dimen@}
\def\y@spacer{\setbox\z@\hbox{$\vcenter{\vskip\dimen@}$}%
  \vrule \@height\ht\z@ \@depth\dp\z@ \@width\z@ \egroup}

\def\spacing@names{
  \normalbaselineskip
  \normallineskip
  \normallineskiplimit
  \footnotebaselineskip
  \interfootnoteskip
  \parskip
  \refbetweenskip
  \abovedisplayskip
  \belowdisplayskip
  \abovedisplayshortskip
  \belowdisplayshortskip
  \chskiptamount
  \chskiplamount
  \secskipamount
  }

\def\twelve@spacing{
  14\p@              +5\p@        *
  \p@                +\z@         *
  \z@                +\z@         *
  14\p@              +\p@         *
  20\p@              +\z@         *
  5\p@\@plus\p@      +-2\p@       *
  \z@                +6\p@        *
  8\p@\@plus2\p@\@minus3\p@  +%
    4\p@\@plus3\p@\@minus5\p@     *
  8\p@\@plus2\p@\@minus3\p@  +%
    4\p@\@plus3\p@\@minus5\p@     *
  \p@\@plus2\p@\@minus\p@    +%
    4\p@\@plus3\p@\@minus2\p@     *
  8\p@\@plus2\p@\@minus3\p@  +%
    \p@\@plus2\p@\@minus2\p@      *
  20\p@\@plus5\p@    +\z@         *
  5.5\p@             +\z@         *
  6\p@\@plus2\p@     +\z@         *
  }

\def\ten@spacing{
  11\p@              +4.5\p@      *
  \p@                +\z@         *
  \z@                +\z@         *
  12\p@              +\p@         *
  16\p@              +\z@         *
  5\p@\@plus\p@      +-2\p@       *
  \z@                +5\p@        *
  8\p@\@plus2\p@\@minus3\p@  +%
    4\p@\@plus3\p@\@minus5\p@     *
  8\p@\@plus2\p@\@minus3\p@  +%
    4\p@\@plus3\p@\@minus5\p@     *
  \p@\@plus2\p@\@minus\p@    +%
    4\p@\@plus3\p@\@minus2\p@     *
  8\p@\@plus2\p@\@minus3\p@  +%
    \p@\@plus2\p@\@minus2\p@      *
  20\p@\@plus5\p@    +\z@         *
  5.5\p@             +\z@         *
  6\p@\@plus2\p@     +\z@         *
  }

\def\eight@spacing{
  9\p@               +3.5\p@      *
  \p@                +\z@         *
  \z@                +\z@         *
  10\p@              +\p@         *
  14\p@              +\z@         *
  5\p@\@plus\p@      +-2\p@       *
  \z@                +5\p@        *
  8\p@\@plus2\p@\@minus3\p@  +%
    4\p@\@plus3\p@\@minus5\p@     *
  8\p@\@plus2\p@\@minus3\p@  +%
    4\p@\@plus3\p@\@minus5\p@     *
  \p@\@plus2\p@\@minus\p@    +%
    4\p@\@plus3\p@\@minus2\p@     *
  8\p@\@plus2\p@\@minus3\p@  +%
    \p@\@plus2\p@\@minus2\p@      *
  20\p@\@plus5\p@    +\z@         *
  5.5\p@             +\z@         *
  6\p@\@plus2\p@     +\z@         *
  }

\twelvepoint


\def\large{\par \bgroup \twelvepoint \after@arg\@size}
\def\medium{\par \bgroup \tenpoint \after@arg\@size}
\def\small{\par \bgroup \eightpoint \after@arg\@size}
\def\@size{\par \egroup}

\def\LARGE#1{{\twelve@point #1}}
\def\MEDIUM#1{{\ten@point #1}}
\def\SMALL#1{{\eight@point #1}}

\def\submittextone{Zur Ver\"offentlichung in\else Submitted to}
\def\submittexttwo{ eingereicht\else}
\def\abstracthead{Zusammenfassung\else Abstract}
\def\ackhead{Danksagung\else Acknowledgements}
\def\appendixhead{Anhang\else Appendix}
\def\eqabbr{Gl\else eq}
\def\eqsabbr{Gln\else eqs}
\def\figpref{Abb\else Fig}
\def\fighead{Abbildungen\else Figure captions}
\def\figabbr{Bild\nodot\else Fig}
\def\tabpref{Tab\else Tab}
\def\tabhead{Tabellen\else Table captions}
\let\tababbr=\tabpref
\def\refpref{Lit\else Ref}
\def\refhead{Literaturverzeichnis\else References}
\def\refabbr{???\else Ref}
\def\refsabbr{????\else Refs}
\def\tocpref{Inh\else Toc}
\def\tochead{Inhaltsverzeichnis\else Table of contents}
\def\footpref{Anm\else Foot}
\def\foothead{Anmerkungen\else Footnotes}
\def\prfhead{Beweis\else Proof}


\def\UPPERCASE#1{\edef\n@xt{#1}\uppercase\expandafter{\n@xt}}

\let\headlinestyle=\twelverm
\let\footlinestyle=\twelverm
\let\pagestyle=\twelverm       
\let\titlestyle=\bf            
\let\authorstyle=\caps
\let\addressstyle=\sl
\let\sectstyle=\caps           

\let\headstyle=\UPPERCASE      
\let\captionstyle=\it          
\let\journalstyle=\sl
\let\volumestyle=\bf



\def\namrefindent{2em}


\let\footstyle=\empty          

\let\stmttitlestyle=\bf        
\let\stmtstyle=\sl
\let\prftitlestyle=\caps
\let\prfstyle=\sl


\def\skipuserexit{\setbox\z@\box\@cclv}  
\def\shipuserexit{\unvbox\@cclv}         

\def\chapuserexit{\sectuserexit}         
\def\appuserexit{\chapuserexit}          
\def\sectuserexit{\secsuserexit}         
\let\secsuserexit=\relax                 

\newcount\firstp@ge   \firstp@ge=-10000
\newcount\lastp@ge   \lastp@ge=10000
\outer\def\pagesel#1#2{\global\firstp@ge#1 \global\lastp@ge#2
  \wlog{**************************}\wlog{*}%
  \wlog{* don't use \string\p agesel}%
  \wlog{* this will not be supported in future}%
  \wlog{*}\wlog{**************************}%
  \wlog{(* pages #1-#2 selected for printing, others will be skipped *)}}

\newbox\pageb@x

\outer\def\toppagenum{\glet\page@tbn T%
  \glet\headb@x\pageb@x \glet\footb@x\voidb@x}
\outer\def\botpagenum{\glet\page@tbn B%
  \glet\headb@x\voidb@x \glet\footb@x\pageb@x}
\outer\def\nopagenum{\glet\page@tbn N%
  \glet\headb@x\voidb@x \glet\footb@x\voidb@x}

\outer\def\lefthead{\glet\head@lrac L}
\outer\def\righthead{\glet\head@lrac R}
\outer\def\althead{\glet\head@lrac A}
\outer\def\centhead{\glet\head@lrac C}
\outer\def\leftfoot{\glet\foot@lrac L}
\outer\def\rightfoot{\glet\foot@lrac R}
\outer\def\altfoot{\glet\foot@lrac A}
\outer\def\centfoot{\glet\foot@lrac C}

\newtoks\lheadtext   \newtoks\cheadtext   \newtoks\rheadtext
\newtoks\lfoottext   \newtoks\cfoottext   \newtoks\rfoottext

\headline={\headlinestyle \head@foot\skip@head\head@lrac
  \lheadtext\cheadtext\rheadtext\headb@x}
\footline={\footlinestyle \head@foot\skip@foot\foot@lrac
  \lfoottext\cfoottext\rfoottext\footb@x}
\lheadtext={}   \cheadtext={}   \rheadtext={}
\lfoottext={}   \cfoottext={}   \rfoottext={}

\newbox\page@strut
\setbox\page@strut\hbox{\vrule \@height 15mm\@depth 10mm\@width \z@}

\def\head@foot#1#2#3#4#5#6{\unhcopy\page@strut
  \if#1T\hfil \else
    \if#2C\head@@foot{\the#3}{\copy#6}{\the#5}\else
      \if#2A\ifodd\pageno \let#2R\else \let#2L\fi \fi
      \if#2R\head@@foot{\the#4}{\the#5}{\copy#6}\else
        \head@@foot{\copy#6}{\the#3}{\the#4}\fi \fi \fi}
\def\head@@foot#1#2#3{\plainrlap{#1}\hfil#2\hfil\plainllap{#3}}

\let\startpage=\relax  

\outer\def\pageall{\glet\page@ac A%
  \global\countdef\pageno\z@ \global\pageno\@ne
  \global\countdef\pageno@pref\@ne \gz@\pageno@pref
  \glet\page@pref\empty \glet\page@reset\count@
  \glet\chap@break\chskipt \outer\gdef\startpage{\global\pageno}}
\outer\def\pagechap{\glet\page@ac C%
  \global\countdef\pageno\@ne \gz@\pageno
  \global\countdef\pageno@pref\z@ \gz@\pageno@pref
  \gdef\page@pref{\dash@pref}%
  \gdef\page@reset{\global\pageno\@ne \global\pageno@pref}%
  \glet\chap@break\superendpage
  \outer\gdef\startpage##1.{\global\pageno@pref##1\global\pageno}}


\hsize=15 cm   \hoffset=0 mm
\vsize=22 cm   \voffset=0 mm

\newdimen\hoffset@corr@p   \newdimen\voffset@corr@p
\newdimen\hoffset@corrm@p   \newdimen\voffset@corrm@p
\newdimen\hoffset@corr@l   \newdimen\voffset@corr@l
\newdimen\hoffset@corrm@l   \newdimen\voffset@corrm@l

\outer\def\portrait{\switch@pl P%
  \glet\hoffset@corr\hoffset@corr@p
  \glet\voffset@corr\voffset@corr@p
  \glet\hoffset@corrm\hoffset@corrm@p
  \glet\voffset@corrm\voffset@corrm@p}
\outer\def\landscape{\switch@pl L%
  \glet\hoffset@corr\hoffset@corr@l
  \glet\voffset@corr\voffset@corr@l
  \glet\hoffset@corrm\hoffset@corrm@l
  \glet\voffset@corrm\voffset@corrm@l}
\def\switch@pl#1{\if #1\ori@pl \else \superendpage \glet\ori@pl#1%
  \dimen@\ht\page@strut \advance\dimen@\dp\page@strut
  \advance\vsize\dimen@ \dimen@ii\hsize \global\hsize\vsize
  \advance\dimen@ii-\dimen@ \global\vsize\dimen@ii \fi}
\let\ori@pl=P

\def\m@g{\dimen@\ht\page@strut \advance\dimen@\dp\page@strut
  \advance\vsize\dimen@ \divide\vsize\count@
  \multiply\vsize\mag \advance\vsize-\dimen@
  \divide\hsize\count@ \multiply\hsize\mag
  \divide\dimen\footins\count@ \multiply\dimen\footins\mag
  \mag\count@}

\output={\physoutput}

\def\physoutput{\make@lbl
  \ifnum \pageno<\firstp@ge \skipp@ge \else
  \ifnum \pageno>\lastp@ge \skipp@ge \else \shipp@ge \fi \fi
  \advancepageno \skippagenum F\skipheadline F\skipfootline F%
  \ifnum\outputpenalty>-\@MM \else \dosupereject \fi}

\def\skippagenum{\glet\skip@page}
\def\skipheadline{\glet\skip@head}
\def\skipfootline{\glet\skip@foot}

\def\skipp@ge{{\skipuserexit \setbox\z@\box\topins
  \setbox\z@\box\footins}\deadcycles\z@}
\def\shipp@ge{\setbox\pageb@x\hbox{%
    \if F\skip@page \pagestyle{\page@pref \folio}\fi}%
  \dimen@-.5\hsize \advance\dimen@\hoffset@corrm
  \divide\dimen@\@m \multiply\dimen@\mag
  \advance\hoffset\dimen@ \advance\hoffset\hoffset@corr
  \dimen@\ht\page@strut \advance\dimen@\dp\page@strut
  \advance\dimen@\vsize \dimen@-.5\dimen@
  \advance\dimen@\voffset@corrm
  \divide\dimen@\@m \multiply\dimen@\mag
  \advance\voffset\dimen@ \advance\voffset\voffset@corr
  \shipout\vbox{\makeheadline \vbadness\@M \setbox\z@\pagebody
    \dimen@\dp\z@ \box\z@ \kern-\dimen@ \makefootline}}

\def\pagecontents{\ifvbox\topins\unvbox\topins\fi
  \dimen@\dp\@cclv \shipuserexit 
  \ifvbox\footins 
    \vskip\skip\footins \footnoterule \unvbox\footins\fi
  \ifr@ggedbottom \kern-\dimen@ \vfil \fi}

\def\folio{\ifnum\pageno<\z@ \ifcase\langu@ge \MEDIUM{\uppercase
  \expandafter{\romannumeral-\pageno}}\else \romannumeral-\pageno \fi
  \else \number\pageno \fi}

\def\makeheadline{\line{\the\headline}\nointerlineskip}
\def\makefootline{\nointerlineskip \line{\the\footline}}

\skippagenum=F   \skipheadline=F   \skipfootline=F

\outer\def\titlepage{\glet\titl@fill\vfil}
\outer\def\notitlepage{\gdef\titl@fill{\vskip20\p@}}

\newbox\t@pleft   \newbox\t@pright
\def\t@pinit{%
  \global\setbox\t@pleft\vbox{\hrule \@height\z@ \@width.26\hsize}%
  \global\setbox\t@pright\copy\t@pleft}
\t@pinit

\def\topleft{\t@p\t@pleft}
\def\topright{\t@p\t@pright}
\def\t@p#1#2{\global\setbox#1\vtop{\unvbox#1\hbox{\strut #2}}}

\outer\def\submit#1{\topleft{\case@language\submittextone}%
  \topleft{{#1}\case@language\submittexttwo}}

\let\pubdate=\topright

\outer\def\title{\vbox{\line{\box\t@pleft \hss \box\t@pright}}%
  \skippagenum T\skipheadline T\skipfootline T%
  \t@pinit \titl@fill \vskip\chskiptamount \@title}
\let\titcon=\relax  
\outer\def\titcon{\errmessage{please use \noexpand\nl in the title
  instead of \noexpand\titcon}\unchskip \@title}
\def\@title#1{\ctrlines\titlestyle{#1}\chskipl}
\def\titl@#1{\edef\n@xt{\noexpand\@title{#1}}\n@xt}

\def\author{\aut@add\authorstyle}
\def\autcon{\and@con \author}
\def\address{\aut@add\addressstyle}
\def\addcon{\and@con \address}
\def\and@con{\titl@fill \ctrline{\case@language{und\else and}}}

\def\aut@add#1{\titl@fill \ctrlines{#1\use@nl}}
\def\use@nl{\let\\\use@@nl}
\def\use@@nl{\errmessage{please use \noexpand\nl in addresses and
  (lists of) authors instead of \string\\}\nl}

\def\abstract{\titl@fill \he@d{\case@language\abstracthead}%
  \after@arg\titl@fill}

\def\ack{\chskipt \he@d{\case@language\ackhead}}

\def\he@d#1{\ctrline{\headstyle{#1}}\chskipl}

\message{chapters, sections and appendices,}


\newtoks\l@names   \l@names={\\\the@label}
\let\the@label=\empty

\def\label{\num@lett\@label}
\def\@label#1{\def@name\l@names#1{\the@label}}
\def\quote{\num@lett\empty}

\newinsert\lbl@ins
\count\lbl@ins=0   \dimen\lbl@ins=\maxdimen   \skip\lbl@ins=0pt
\newcount\lbln@m   \lbln@m=0
\let\lbl@saved\empty

\def\pagelabel{\num@lett\@pagelabel}
\def\@pagelabel#1{\ifx#1\undefined \let#1\empty \fi
  \toks@\expandafter{#1}\expandafter\testcr@ss\the\toks@\cr@ss\@@
  \ifcr@ss\else
    \toks@{\cr@ss\lbl@undef}\def@name\l@names#1{\the\toks@}\fi
  \g@ne\lbln@m \insert\lbl@ins{\vbox{\vskip\the\lbln@m sp}}%
  \count@\lbln@m \do@label\store@label#1}
\begingroup \let\save=\relax  
  \gdef\lbl@undef{\message{unresolved \string\pagelabel, use
    \string\save\space and \string\crossrestore}??}
\endgroup
\def\do@label#1{\expandafter#1\csname\the\count@\endcsname}
\def\store@label#1#2{\expandafter\gdef\expandafter\lbl@saved
  \expandafter{\lbl@saved#1#2}}

\def\make@lbl{\setbox\z@\vbox{\let\MEDIUM\relax
  \unvbox\lbl@ins \loop \setbox\z@\lastbox \ifvbox\z@
    \count@\ht\z@ \do@label\make@label \repeat}}
\def\make@label#1{\def\make@@label##1#1##2##3#1##4\@@{%
    \gdef\lbl@saved{##1##3}%
    \ifx @##4@\errmessage{This can't happen}\else
    \def@name\l@names##2{\page@pref\folio}\fi}%
  \expandafter\make@@label\lbl@saved#1#1#1\@@}

\outer\def\lblrestore{\all@restore\l@names}


\let\sect=\relax  \let\s@ct=\relax  

\def\chapinit{\chap@init{\chap@pref}\glet\sect@@eq\@chap@sect@eq
  \sectinit}
\def\appinit{\chap@init{\char\the\appn@m}\glet\sect@@eq\@chap@eq
  \glet\sect@dot@pref\empty \glet\sect@pref\dot@pref
  \glet\sect\undefined}
\def\sectinit{\xdef\sect@dot@pref{\the\sectn@m.}%
  \xdef\sect@pref{\dot@pref\sect@dot@pref}\glet\sect\s@ct}
\def\chap@init#1{\xdef\the@label{#1}\xdef\dot@pref{\the@label.}%
  \xdef\dash@pref{\the@label--}\glet\chap@@eq\@chap@eq}


\newcount\chapn@m   \chapn@m=0

\outer\def\chappage{\glet\chap@page T}
\outer\def\nochappage{\glet\chap@page F}

\outer\def\arabicchapnum{\glet\chap@ar A\gdef\chap@pref{\the\chapn@m}}
\outer\def\romanchapnum{\glet\chap@ar R%
  \gdef\chap@pref{\uppercase{\romannumeral\chapn@m}}}

\let\chap=\relax  \let\ch@p=\relax  

\outer\def\chapters{\glet\chap@yn Y\glet\chap\ch@p
  \chap@init{0}\glet\sect@@eq\@chap@sect@eq \sectinit}
\outer\def\nochapters{\glet\chap@yn N\glet\chap\undefined
  \glet\dot@pref\empty \glet\dash@pref\empty
  \glet\chap@@eq\@eq \glet\sect@@eq\@sect@eq \sectinit}

\outer\def\ch@p#1{\if T\chap@page \superendpage \else \chap@break \fi
  \g@ne\chapn@m \sect@reset \chapinit \page@reset\chapn@m
  \eq@reset \fig@reset \tab@reset
  \toks@{\dot@pref}\toks@ii{#1}\chapuserexit
  \titl@{\the\toks@\ \the\toks@ii}%
  \ifnum\auto@toc>\m@ne \toks@store{#1}\@toc\dot@pref \fi}



\def\sec@title#1#2#3#4#5#6#{\ifx @#6@\g@ne#1\else\global#1#6\fi
  #2\secskipt \xdef\the@label{#3\the#1}\xdef#4{\the@label.}%
  \read@store{\sec@@title#4#5}}
\def\sec@@title#1#2#3#4{\toks@{\the@label.}\toks@ii\toks@store
  #4\bpargroup #2\varitem{\the\toks@}\interlinepenalty\@M
    \let\nl\lb \the\toks@ii \par\nobreak
  \ifnum#3<\auto@toc \@toc{#1}\fi}

\newcount\sectn@m   \sectn@m=0
\def\sect@reset{\gz@\sectn@m}
\outer\def\s@ct{\sec@title\sectn@m
  {\secs@reset \sectinit \eq@@reset \fig@@reset \tab@@reset}%
  \dot@pref\sect@pref{\sectstyle\z@\sectuserexit}}

\let\secs@reset=\relax


\def\sect@lev{\@ne}         
\def\sect@id{sect}          
\def\secs@id{secs}          

\outer\def\newsect{\begingroup \count@\sect@lev
  \let\@\endcsname \let\or\relax
  \edef\n@xt{\new@sect}\advance\count@\@ne
  \xdef\sect@lev{\the\count@\space}\n@xt
  \glet\sect@id\secs@id \xdef\secs@id{\secs@id s}\endgroup}
\begingroup \let\newcount=\relax
  \gdef\new@sect{\wlog{\noexpand\string\secs@nm\@= subsection
      level \noexpand\sect@lev}\noexpand\newcount\secs@nm\n@m@
    \gdef\secs@nm\@reset@{\noexpand\gz@\secs@nm\n@m@}%
    \outer\gdef\secs@nm\@{\noexpand\sec@title\secs@nm\n@m@
      \secs@nm s\@reset@ \csn@me\sect@id\@pref@ \secs@nm\@pref@
      {\secs@nm\style@{\the\count@}\secs@nm\userexit@}}%
    \glet\secs@nm s\@reset@ \relax
    \gdef\secs@nm\style@{\csn@me\sect@id\style@}%
    \gdef\secs@nm\userexit@{\secs@nm s\userexit@}%
    \glet\secs@nm s\userexit@\relax
    \outer\xdef\csn@me toc\secs@id\@
      {\global\auto@toc\noexpand\the\count@\space}%
    \xdef\noexpand\save@@toc{\save@@toc\or\secs@id}}
\endgroup

\def\n@m@{n@m\@}
\def\@pref@{@pref\@}
\def\@reset@{@reset\@}
\def\style@{style\@}
\def\userexit@{userexit\@}

\def\secs@nm{\csn@me\secs@id}
\def\csn@me{\expandafter\noexpand\csname}


\newcount\appn@m   \appn@m=64

\outer\def\appendix{\if T\chap@page \superendpage \else\chap@break \fi
  \g@ne\appn@m \secs@reset \appinit \page@reset\appn@m
  \eq@reset \fig@reset \tab@reset \futurelet\n@xt \app@ndix}

\def\app@ndix{\ifcat\bgroup\noexpand\n@xt \expandafter\@ppendix \else
  \expandafter\@ppendix\expandafter\unskip \fi}

\def\@ppendix#1{\toks@{\case@language\appendixhead^^>\dot@pref}
  \toks@ii{#1}\appuserexit
  \titl@{\the\toks@\ \the\toks@ii}%
  \ifnum\auto@toc>\m@ne \toks@store{#1}%
  \@toc{\case@language\appendixhead\ \dot@pref}\fi}

\def\app#1{\chskipt \he@d{\case@language\appendixhead\ #1}}

\def\num@lett{\cat@lett \num@@lett}
\def\num@@lett#1#2{\egroup #1{#2}}

\def\num@l@tt{\cat@lett \num@@l@tt}
\def\num@@l@tt#1#{\egroup #1}

\def\cat@lett{\bgroup
  \catcode`\0\l@tter \catcode`\1\l@tter \catcode`\2\l@tter
  \catcode`\3\l@tter \catcode`\4\l@tter \catcode`\5\l@tter
  \catcode`\6\l@tter \catcode`\7\l@tter \catcode`\8\l@tter
  \catcode`\9\l@tter \catcode`\'\l@tter}

\def\quote@all#1{\leavevmode\hbox{\mathcode`\-\dq 707B$#1$}}
\def\use{\num@lett\@use}
\def\@use{\setbox\z@\hbox}


\newtoks\e@names   \e@names={}
\newcount\eqn@m   \eqn@m=0

\outer\def\equall{\glet\eq@acs A\glet\eq@pref\empty
  \glet\def@eq\@eq \glet\eq@reset\relax \glet\eq@@reset\relax}
\outer\def\equchap{\glet\eq@acs C\gdef\eq@pref{\dot@pref}%
  \gdef\def@eq{\chap@@eq}\glet\eq@reset\eqz@ \glet\eq@@reset\relax}
\outer\def\equsect{\glet\eq@acs S\gdef\eq@pref{\sect@pref}%
  \gdef\def@eq{\sect@@eq}\glet\eq@reset\eqz@ \glet\eq@@reset\eqz@}
\def\eqz@{\gz@\eqn@m}

\outer\def\equfull{\glet\eq@fs F\gdef\eq@@fs{\let\test@eq\full@eq}}
\outer\def\equshort{\glet\eq@fs S\glet\eq@@fs\relax}

\def\@eq(#1){#1}
\def\@chap@eq{\noexpand\chap@eq\@eq}
\def\@sect@eq{\noexpand\sect@eq\@eq}
\def\@chap@sect@eq{\noexpand\chap@sect@eq\@eq}

\def\chap@eq{\test@eq\empty\dot@pref}
\def\sect@eq{\test@eq\empty\sect@dot@pref}
\def\chap@sect@eq{\test@eq\sect@eq\dot@pref}
\def\test@eq#1#2#3.{\def\n@xt{#3.}\ifx#2\n@xt \let\n@xt#1\fi \n@xt}
\def\full@eq#1#2{}
\def\short@eq#1#2#3.{#1}

\outer\def\equleft{\glet\eq@lrn L\glet\eqtag\leqno
  \glet\eq@tag\leq@no}
\outer\def\equright{\glet\eq@lrn R\glet\eqtag\eqno
  \glet\eq@tag\eq@no}
\outer\def\equnone{\glet\eq@lrn N\glet\eqtag\n@eqno
  \glet\eq@tag\neq@no}

\begingroup
  \catcode`\$=\active \catcode`\*=3 \lccode`\*=`\$
  \lowercase{\gdef\n@eqno{\catcode`\$\active
                \def$${\egroup **}\setbox\z@\hbox\bgroup}}
\endgroup
\def\eq@no{\llap{$\@lign##$}\tabskip\z@skip}
\def\leq@no{\kern-\displaywidth \rlap{$\@lign##$}\tabskip\displaywidth}
\def\neq@no{\@use{$\@lign##$}\tabskip\z@skip}

\def\displaylines{\afterassignment\display@lines \@eat}
\def\display@lines{\displ@y
  \halign\n@xt\hbox to\displaywidth{$\@lign\hfil\displaystyle##\hfil$}%
    &\span\eq@tag\crcr}

\def\eqalignno{\let\eq@@tag\eq@no \eqalign@tag}
\def\leqalignno{\let\eq@@tag\leq@no \eqalign@tag}
\def\eqaligntag{\let\eq@@tag\eq@tag \eqalign@tag}
\def\eqalign@tag{\afterassignment\eqalign@@tag \@eat}
\def\eqalign@@tag{\displ@y
  \tabskip\centering \halign to\displaywidth\n@xt
    \hfil$\@lign\displaystyle{##}$\tabskip\z@skip
    &$\@lign\displaystyle{{}##}$\hfil\tabskip\centering
    &\span\eq@@tag\crcr}

\def\fulltag#1{{\let\test@eq\full@eq#1}}
\def\shorttag#1{{\let\test@eq\short@eq#1}}

\def\eq{\g@ne\eqn@m \make@eq\empty}
\def\make@eq#1{(\eq@pref\the\eqn@m #1)}
\def\EQ{\eq \num@lett\eq@save}
\def\eq@save#1{\def@name\e@names#1{\expandafter\def@eq\make@eq\empty}}

\def\eqn{\eqtag\eq}
\def\EQN{\eqtag\EQ}

\def\eqadv{\g@ne\eqn@m}
\def\EQADV{\eqadv \num@lett\eq@save}

\newcount\seqn@m   \seqn@m=96

\def\subeqbegin{\global\seqn@m96 \subeq}
\def\SUBEQBEGIN{\global\seqn@m96 \SUBEQ}
\def\subeq{\g@ne\seqn@m \make@eq{\char\seqn@m}}
\def\SUBEQ{\num@lett\@SUBEQ}
\def\@SUBEQ#1{\subeq \def@name\e@names#1{\char\the\seqn@m}}

\def\eqapp{\num@lett\@eqapp}
\def\@eqapp#1#2{(\fulltag#1#2)}

\def\queq{\num@lett\@queq}
\def\@queq#1{\quote@all{\eq@@fs(#1)}}
\def\qeq{\case@abbr\eqabbr\queq}
\def\qeqs{\case@abbr\eqsabbr\queq}

\outer\def\eqrestore{\all@restore\e@names}


\newtoks\toks@store
\newtoks\file@list \file@list={1234567}
\def\@sysut{\jobname.sysut}
\let\ext@ft=F

\begingroup
  \let\storebox=\relax \let\refnam=\relax  
  \let\RFfile=\relax \let\RFext=\relax     
  \newhelp\opt@help{The options \string\refnam, \string\RFfile\space
    and \string\RFext\space are incompatible with \string\storebox.
    Your request will be ignored.}
  \global\opt@help=\opt@help 
  \gdef\opt@err{{\errhelp\opt@help \errmessage{Incompatible options}}}
\endgroup

\begingroup
  \let\storebox=\relax \let\storelist=\relax  
  \let\storefile=\relax \let\RFfile=\relax    
  \gdef\case@store{%
    \glet\storebox\undefined
    \if B\store@blf \glet\storebox\empty \glet\case@store\case@box
      \else \glet\box@store\undefined
      \glet\box@out\undefined \glet\box@print\undefined
      \glet\box@save\undefined \glet\box@kill\undefined \fi
    \glet\storelist\undefined
    \if L\store@blf \glet\storelist\empty \glet\case@store\case@list
      \else \glet\list@store\undefined
      \glet\list@out\undefined \glet\list@print\undefined
      \glet\list@save\undefined \glet\list@kill\undefined \fi
    \glet\storefile\undefined
    \if F\store@blf \glet\storefile\empty \glet\case@store\case@file
      \else \store@setup
      \glet\file@out\undefined \glet\file@print\undefined
      \glet\filef@rm@t\undefined \glet\fil@f@rm@t\undefined
      \glet\file@save\undefined \glet\file@kill\undefined \fi
    \glet\case@box\undefined \glet\case@list\undefined
    \glet\case@file\undefined \glet\store@setup\undefined
    \case@store}
  \gdef\store@setup{\ifx \RFfile\undefined \glet\file@store\undefined
    \glet\file@open\undefined \glet\file@close\undefined
    \glet\file@wlog\undefined \glet\file@free\undefined
    \glet\file@copy\undefined \glet\file@read\undefined \fi}
\endgroup

\outer\def\storebox{\if T\ext@ft \opt@err \else
    \if L\RF@lfe \if N\ref@sbn \opt@err
    \else \glet\store@blf B\fi \else \opt@err \fi \fi}
\outer\def\storelist{\glet\store@blf L}
\outer\def\storefile{\glet\store@blf F}

\def\read@store{\bgroup \@read@store}
\def\read@@store{\bgroup \catcode`\@\l@tter \@read@store}
\def\@read@store#1{\def\after@read{\egroup \toks@store\toks@i
    #1\after@read \ignorespaces}%
  \catcode`\^^M\active \afterassignment\after@read \global\toks@i}
\def\afterread#1{\bgroup \def\after@read{\egroup #1\after@read}}
\let\after@read=\relax
\begingroup
  \catcode`\:=\active
  \gdef\write@save#1{\write@store{:restore\@type{#1}}}
\endgroup
\def\write@store#1{\s@ve{#1{\the\toks@store}}}

\def\f@rm@t#1#2{\bgroup \ignorefoot
  \leftskip\z@skip \rightskip\z@skip \f@rmat
  \ifx @#1@\everypar{\b@format}\else
    \varitem\@indent{#1}\b@format \fi #2\e@format}
\let\f@rmat=\nointerlineskip
\def\form@t{\unskip \strut \par \@break \eform@t}
\def\b@format{\glet\e@format\form@t \strut}
\def\eform@t{\egroup \glet\e@format\eform@t}
\let\e@format=\eform@t

\def\@store{\case@store\box@store\list@store\file@store}
\def\@sstore#1#2#3{\par \noindent \bgroup \captionstyle
  \case@abbr#2#3:\enskip \the\toks@store \egroup \par \@store#1{#3.}}
\def\@add#1{\read@store{\@store#1{}}}
\def\@out#1#2#3#4#5{\case@store\box@out\list@out\file@out#1\begingroup
  \if T#2\let\chap@break\superendpage \fi \chap@break
  \chap@init{\case@language#3}%
  \if C\page@ac \skippagenum T\fi
  \page@reset6#1%
  \@style \he@d{\strut\case@language#4}\@break \@print#1%
  \ifx\chap@break\superendpage \superendpage \fi
  \def\\##1{\glet##1\undefined}\the#5\global#5\emptyt@ks
  \endgroup \fi}
\def\@print{\case@store\box@print\list@print\file@print}
\def\@save{\case@store\box@save\list@save\file@save}
\def\@kill{\case@store\box@kill\list@kill\file@kill}
\def\@ext#1#2#3 {\if B\store@blf \opt@err \else
  \glet\ext@ft T\@add#1{#2#3 }\bgroup
   \def\@store##1##2{}\input#3 \egroup \fi}
\def\@@ext#1#2{\let#1#2\everypar\emptyt@ks
  \def\read@store##1{\relax##1}\def\@store##1{\f@rm@t}\input}

\def\case@box#1#2#3{\case@@store#1%
  \fig@box\tab@box\ref@box\toc@box\foot@box}
\def\case@list#1#2#3{\case@@store#2%
  \fig@list\tab@list\ref@list\toc@list\foot@list}
\def\case@file#1#2#3{\case@@store{\expandafter#3}%
  \fig@file\tab@file\ref@file\toc@file\foot@file}

\def\case@@store#1#2#3#4#5#6#7{\ifcase#7%
  \toks@{\fig@type#1#2}\or
  \toks@{\tab@type#1#3}\or
  \toks@{\ref@type#1#4}\or
  \toks@{\toc@type#1#5}\or
  \toks@{\foot@type#1#6}\fi
  \expandafter\let\expandafter\@type\the\toks@}
\def\@style{\csname\@type style\endcsname}
\def\@indent{\csname\@type indent\endcsname}
\def\@break{\csname\@type break\endcsname}

\def\box@store#1#2{\global\setbox#1\vbox
  {\ifvbox#1\unvbox#1\fi \@style \f@rm@t{#2}{\the\toks@store}}}
\def\box@out{\ifvbox}
\def\box@print{\vskip\baselineskip \unvbox}
\begingroup
  \catcode`\:=\active \catcode`\;=\active
  \gdef\box@save#1{\wlog{; Unable to save text for
    \@type's with option :storebox}}
\endgroup
\def\box@kill#1{{\setbox\z@\box#1}}

\def\list@store#1#2{\toks@\expandafter{#1\\}%
  \xdef#1{\the\toks@ {#2}{\the\toks@store}}}
\def\list@out#1{\ifx #1\empty \else}
\def\list@print#1{\let\\\f@rm@t #1\glet#1\empty}
\def\list@save{\def\\##1##2{\toks@store{##2}\write@save{##1}}%
  \newlinechar`\^^M}
\def\list@kill#1{\glet#1\empty}

\begingroup
  \catcode`\:=\active
  \gdef\file@store#1#2#3#4{%
    \if0#3\expandafter\file@open\the\file@list\@@#1#2\fi
    {\newlinechar`\^^M\let\save@write#2\write@store{::{#4}}}}
\endgroup
\def\file@open#1#2\@@#3#4{\immediate\openout#4\@sysut#1
  \gdef#3{#3#4#1}\global\file@list{#2}\file@wlog{open}#1}
\def\file@wlog#1#2{\wlog{#1 \@sysut#2 for \@type's}}
\def\file@out#1#2#3{\if0#3\else}
\def\file@print#1#2#3{\file@close#2#3\let\\\filef@rm@t
  \file@copy#1#2#3}
\def\filef@rm@t#1{\bgroup \catcode`\@\l@tter \fil@f@rm@t{#1}}
\def\fil@f@rm@t#1#2{\egroup \f@rm@t{#1}{#2}}
\def\file@save#1#2#3{\if0#3\else \file@close#2#3%
  \def\\##1{\read@@store{\expandafter\file@store#1{##1}%
    \write@save{##1}}}%
  \newlinechar`\^^M\file@copy#1#2#3\fi}
\def\file@kill#1#2#3{\if0#3\else \file@close#2#3\file@free#1#2#3\fi}
\def\file@close#1#2{\immediate\closeout#1\file@wlog{close}#2}
\def\file@copy#1#2#3{\file@free#1#2#3\file@read#3}
\def\file@read#1{\input\@sysut#1 }
\def\file@free#1#2#3{\gdef#1{#1#20}%
  \global\file@list\expandafter{\the\file@list#3}}

\def\if@t#1#2#3#4#5#6{\glet#1#6\gdef#2{\dot@pref}\glet#3#5\glet#4\relax
  \if#6A\glet#2\empty \glet#3\relax \fi
  \if#6S\gdef#2{\sect@pref}\glet#4#5\fi}
\def\bf@t#1{\let\f@t#1\num@l@tt}
\def\ef@t#1#2#3#4#{\ifx @#4@\g@ne#1\xdef\thef@tn@m{#2\the#1}\else
  \gdef\thef@tn@m{#4}\fi #3\read@store\f@t}
\def\@extf@t#1#2#{\gdef\thef@tn@m{#1}\f@t}


\newtoks\f@names   \f@names={}
\newcount\fign@m   \fign@m=0
\def\fig@type{fig}
\newbox\fig@box
\let\fig@list=\empty
\newwrite\fig@write  \def\fig@file{\fig@file\fig@write0}

\outer\def\figall{\fig@init A}
\outer\def\figchap{\fig@init C}
\outer\def\figsect{\fig@init S}
\def\fig@init{\if@t\fig@acs\fig@pref\fig@reset\fig@@reset\figz@}
\def\figz@{\gz@\fign@m}

\outer\def\figpage{\glet\fig@page T}
\outer\def\nofigpage{\glet\fig@page F}

\def\fig{\bf@t\fig@\@fig}
\def\FIG{\bf@t\fig@\@FIG}
\def\ffig{\bf@t\ffig@\@fig}
\def\FFIG{\bf@t\ffig@\@FIG}

\def\@fig{\ef@t\fign@m\fig@pref\relax}
\def\@FIG#1{\ef@t\fign@m\fig@pref{\def@name\f@names#1{\thef@tn@m}}}

\def\fig@{\@store0{\thef@tn@m .}}
\def\ffig@{\@sstore0\figpref{\thef@tn@m}}
\def\figadd{\@add0}

\def\qufig{\case@abbr\figabbr\num@lett\quote@all}

\outer\def\figout{\@out0\fig@page\figpref\fighead\emptyt@ks}
\outer\def\figkill{\@kill0}
\outer\def\restorefig#1{\read@store{\@store0{#1}}}
\outer\def\figrestore{\all@restore\f@names}

\outer\def\FIGext{\@ext0\FIG@ext}
\def\FIG@ext{\@@ext\@FIG\@extf@t}

\newdimen\spictskip  \spictskip=2.5pt

\def\pict{\bf@t\pict@\@fig}
\def\PICT{\bf@t\pict@\@FIG}

\def\pict@{\vskip\the\toks@store \bpargroup
  \raggedright \captionstyle \varitem{\qufig{\thef@tn@m\,}:}%
  \let\@spict\spict@ \spacefactor998\ignorespaces}

\def\spict#1{\ifnum\spacefactor=998\else \parvskip\spictskip \fi
  \@spict{#1\enskip}\ignorespaces}
\def\spict@#1{\setbox\z@\hbox{#1}\advance\hangindent\wd\z@
  \box\z@ \let\@spict\llap}


\outer\def\graphics{\glet\graphic\gr@phic}
\outer\def\nographics{\glet\graphic\nogr@phic}

\def\gr@phic#1{\vbox\bgroup \def\gr@@@ph{#1}\bfr@me\gr@ph}
\def\nogr@phic#1{\vbox\bgroup \write\m@ne{Insert plot #1}\bfr@me\fr@me}
\def\frame{\vbox\bgroup \bfr@me\fr@me}
\def\bfr@me#1#2#3#4{\tfr@me\z@\dimen@iv#2\relax 
  \dimen@#3\relax \tfr@me\dimen@\dimen@ii#4\relax 
  \setbox\z@\hbox to\dimen@iv{#1}\ht\z@\dimen@ \dp\z@\dimen@ii
  \box\z@ \egroup}
\def\tfr@me#1#2#3\relax{#2#3\relax \ifdim#2<-#1\errhelp\fr@mehelp
  \errmessage{Invalid box size}#2-#1\fi}
\newhelp\fr@mehelp{The \string\wd\space and \string\ht+\string\dp\space
  of a \string\frame\space or \string\graphic\space \string\box\space
  must not be negative and will be changed to 0pt.}

\def\gr@ph{\lower\dimen@ii\gr@@ph b\hfil
  \raise\dimen@\gr@@ph e}
\def\gr@@ph#1{\hbox{\special{^X\gr@@@ph^A}}}         
\def\gr@@ph#1{\hbox{\special{#1plot GKSM \gr@@@ph}}} 
\def\gr@@ph#1{\hbox{\special{^X#1plot GKSM \gr@@@ph^A}}} 
\def\fr@me{\vrule\fr@@@me
  \bgroup \dimen@ii-\dimen@ \fr@@me\dimen@ii \hfilneg
  \bgroup \dimen@-\dimen@ii \fr@@me\dimen@ \vrule\fr@@@me}
\def\fr@@me#1{\advance#1.4\p@ \leaders \hrule\fr@@@me \hss \egroup}
\def\fr@@@me{\@height\dimen@ \@depth\dimen@ii}

\newtoks\t@names   \t@names={}
\newcount\tabn@m   \tabn@m=0
\def\tab@type{tab}
\newbox\tab@box
\let\tab@list=\empty
\newwrite\tab@write  \def\tab@file{\tab@file\tab@write0}

\outer\def\taball{\tab@init A}
\outer\def\tabchap{\tab@init C}
\outer\def\tabsect{\tab@init S}
\def\tab@init{\if@t\tab@acs\tab@pref\tab@reset\tab@@reset\tabz@}
\def\tabz@{\gz@\tabn@m}

\outer\def\tabpage{\glet\tab@page T}
\outer\def\notabpage{\glet\tab@page F}

\def\tab{\bf@t\tab@\@tab}
\def\TAB{\bf@t\tab@\@TAB}
\def\ttab{\bf@t\ttab@\@tab}
\def\TTAB{\bf@t\ttab@\@TAB}

\def\@tab{\ef@t\tabn@m\tab@pref\relax}
\def\@TAB#1{\ef@t\tabn@m\tab@pref{\def@name\t@names#1{\thef@tn@m}}}

\def\tab@{\@store1{\thef@tn@m .}}
\def\ttab@{\@sstore1\tabpref{\thef@tn@m}}
\def\tabadd{\@add1}

\def\qutab{\case@abbr\tababbr\num@lett\quote@all}

\outer\def\tabout{\@out1\tab@page\tabpref\tabhead\emptyt@ks}
\outer\def\tabkill{\@kill1}
\outer\def\restoretab#1{\read@store{\@store1{#1}}}
\outer\def\tabrestore{\all@restore\t@names}

\outer\def\TABext{\@ext1\TAB@ext}
\def\TAB@ext{\@@ext\@TAB\@extf@t}

\newskip\htabskip   \htabskip=1em plus 2em minus .5em
\newdimen\vtabskip  \vtabskip=2.5pt
\newbox\tab@top   \newbox\tab@bot

\let\@hrule=\hrule
\let\@halign=\halign
\let\@valign=\valign
\let\@span=\span
\let\@omit=\omit

\def\@@span{\@span\@omit\@span}
\def\@@@span{\@span\@omit\@@span}

\def\sp@n{\span\@omit\advance\mscount\m@ne} 

\def\table#1#{\vbox\bgroup\offinterlineskip
  \toks@ii{#1\bgroup \unhcopy\tab@top \unhcopy\tab@bot
    ##}\afterassignment\tab@preamble \@eat}


\def\tab@preamble#1\cr{\let\tab@@vrule\tab@repeat
  \let\tab@amp@\empty \let\tab@amp\empty \let\span@\@@span
  \toks@{\tab@space#1&\cr}\the\toks@}
\def\tab@space{\tab@test{ }{}\tab@vrule}                 
\def\tab@vrule{\tab@test\vrule{\tab@add\vrule}
  \tab@@vrule}
\def\tab@repeat{\tab@test&{\tab@add&
    \let\tab@@vrule\tab@template \let\tab@amp@\tab@@amp
    \let\tab@amp&\let\span@\@@@span}\tab@template}
\def\tab@template#1&{\tab@add{\@span\tab@amp@
    \tabskip\htabskip&\tab@setup#1&\tabskip\z@skip##}
  \tab@test\cr{\let\tab@@vrule\tab@exec}\tab@space}      
\def\tab@@amp{&##}

\def\tab@test#1#2#3{\let\tab@comp= #1\toks@{#2}\let\tab@go#3%
  \futurelet\n@xt \tab@@test}
\def\tab@@test{\ifx \tab@comp\n@xt \the\toks@
    \afterassignment\tab@go \expandafter\@eat \else
  \expandafter\tab@go \fi}

\def\tab@add#1{\toks@ii\expandafter{\the\toks@ii#1}}


\def\tab@exec{\tab@r@set\everycr{\tab@body}%
  \def\halign{\tab@r@set \halign}\def\valign{\tab@r@set \valign}%
  \def\omit{\@omit \tab@setup}%
  \def\n@xt##1{\hbox{\dimen@\ht\strutbox\dimen@ii\dp\strutbox
    \advance##1\vtabskip \vrule \@height\dimen@ \@depth\dimen@ii
    \@width\z@}}%
  \setbox\tab@top\n@xt\dimen@ \setbox\tab@bot\n@xt\dimen@ii
  \def\ml##1{\relax                                      
    \ifmmode \let\@ml\empty \else \let\@ml$\fi           
    \@ml\vcenter{\hbox\bgroup\unhcopy\tab@top            
    ##1\unhcopy\tab@bot\egroup}\@ml}%
  \def\nl{\egroup\hbox\bgroup\strut}
  \tabskip\z@skip\@halign\the\toks@ii\cr}

\def\tab@r@set{\let\cr\endline \everycr\emptyt@ks
  \let\halign\@halign \let\valign\@valign
  \let\span\@span \let\omit\@omit}
\def\tab@setup{\relax \iffalse {\fi \let\span\span@ \iffalse }\fi}


\def\tab@body{\noalign\bgroup \tab@@body}           
\def\tab@@body{\futurelet\n@xt \tab@end}            
\def\tab@end{\ifcat\egroup\noexpand\n@xt
    \expandafter\egroup \expandafter\egroup \else        
  \expandafter\tab@blank \fi}
\def\tab@blank{\ifcat\space\noexpand\n@xt
    \afterassignment\tab@@body \expandafter\@eat \else   
  \expandafter\tab@hrule \fi}
\def\tab@hrule{\ifx\hrule\n@xt
    \def\hrule{\@hrule\egroup \tab@body}\else            
  \expandafter\tab@noalign \fi}
\def\tab@noalign{\ifx\noalign\n@xt
    \aftergroup\tab@body \expandafter\@eat@ \else        
  \expandafter\tab@row \fi}
\def\tab@row#1\cr{\toks@ii{\toks@ii{\egroup}
    \tab@item#1&\cr}\the\toks@ii}
\def\tab@item#1&{\tab@add{\tab@amp&#1&}
  \futurelet\n@xt \tab@cr}
\def\tab@cr{\ifx\cr\n@xt
    \expandafter\the\expandafter\toks@ii \else           
  \expandafter\tab@item \fi}

\newtoks\r@names   \r@names={}
\newcount\refn@m   \refn@m=0
\newcount\ref@temp
\def\ref@type{ref}
\newbox\ref@box
\let\ref@list=\empty
\newwrite\ref@write  \def\ref@file{\ref@file\ref@write0}

\begingroup \let\refnam=\relax  
  \newhelp\ref@help{The option \string\refnam\space allows predefined
    references only and is incompatible with \string\qref(s).
    Your request will be ignored.}
  \global\ref@help=\ref@help 
  \gdef\ref@err{{\errhelp\ref@help \errmessage{Invalid request}}}
\endgroup

\begingroup
  \let\refsup=\relax \let\refsqb=\relax  
  \let\refnam=\relax                     
  \gdef\ref@setup{%
    \glet\refsup\undefined
    \if S\ref@sbn \glet\refsup\empty \glet\therefn@m\suprefn@m \fi
    \glet\refsqb\undefined
    \if B\ref@sbn \glet\refsqb\empty \glet\therefn@m\sqbrefn@m \fi
    \glet\refnam\undefined
    \if N\ref@sbn \glet\refnam\empty \glet\the@quref\nam@quref
      \glet\@@ref\ref@err \gdef\qref{\ref@err \quref}\glet\qrefs\qref
      \glet\RF@def@\RF@def@nam
      \glet\RF@find\undefined \glet\RF@search\undefined
      \glet\RF@locate\undefined \glet\@RFread\undefined
      \glet\qurefsup\ref@err \glet\sup@quref\undefined
      \glet\qurefsqb\ref@err \glet\sqb@quref\undefined
      \glet\qurefnum\ref@err \glet\num@quref\undefined
      \glet\ref@restore\ref@err
      \else \gdef\@@ref{\@store2\therefn@m}%
      \gdef\qref{\case@abbr\refabbr\num@lett\quote@all}%
      \gdef\qrefs{\case@abbr\refsabbr\num@lett\quote@all}%
      \glet\RF@def@\RF@def@num \glet\RF@print\undefined
      \gdef\ref@restore{\all@restore\r@names}\fi
    \glet\nam@quref\undefined
    \gdef\quref{\ref@unskip \num@lett\the@quref}%
    \glet\RF@def@num\undefined \glet\RF@def@nam\undefined
    \gdef\RF@restore{\all@restore\R@names}%
    \glet\ref@setup\undefined}
\endgroup

\outer\def\refsup{\glet\ref@sbn S\global\qurefsup}
\outer\def\refsqb{\glet\ref@sbn B\global\qurefsqb}
\outer\def\refnam{\if B\store@blf \opt@err \else \glet\ref@sbn N\fi}

\def\qurefsup{\let\the@quref\sup@quref}
\def\qurefsqb{\let\the@quref\sqb@quref}
\def\qurefnum{\let\the@quref\num@quref}

\outer\def\refpage{\glet\ref@page T}
\outer\def\norefpage{\glet\ref@page F}

\outer\def\refkeep{\glet\ref@kc K}
\outer\def\refclear{\glet\ref@kc C}

\def\ref{\ref@advance \refend \@ref}
\def\REF{\num@lett\@REF}
\def\@REF#1{\ref@name#1\@ref}
\def\refend{\quref{\the\refn@m}}

\def\refs{\ref@advance \ref@temp\refn@m \@ref}
\def\REFS{\num@lett\@REFS}
\def\@REFS#1{\ref@name#1\ref@temp\refn@m \@ref}
\def\refscon{\ref@advance \@ref}

\def\refsend{\quref{\the\ref@temp -\the\refn@m}}

\def\ref@advance{\ref@unskip \g@ne\refn@m}
\def\ref@name{\ref@@name\r@names}
\def\ref@@name#1#2{\ref@advance \def@name#1#2{\the\refn@m}}
\def\ref@unskip{\ifhmode \unskip \fi}

\def\suprefn@m{\the\refn@m .}
\def\sqbrefn@m{$\lbrack \the\refn@m \rbrack$}
\def\@ref{\read@store\@@ref}
\def\@@ref{\ref@setup \@@ref}
\def\refadd{\@add2}

\def\sup@quref#1{\leavevmode \nobreak \quote@all{^{#1}}}
\def\sqb@quref#1{\ \quote@all{\lbrack #1\rbrack}}
\def\num@quref{\ \quote@all}
\def\nam@quref{\@use}
\def\quref{\ref@setup \quref}
\def\qref{\ref@setup \qref}
\def\qrefs{\ref@setup \qrefs}

\outer\def\refout{{\if K\ref@kc \@out2\ref@page\refpref\refhead\emptyt@ks
  \else \@out2\ref@page\refpref\refhead\r@names
  \let\\\@RFdef \the\R@names \global\R@names\emptyt@ks
  \if L\RF@lfe \else \glet\RF@list\empty \gz@\RF@high \fi
  \gz@\refn@m \fi}}
\outer\def\refkill{\@kill2}
\outer\def\restoreref#1{\read@@store{\@store2{#1}}}
\outer\def\refrestore{\ref@restore}
\outer\def\RFrestore{\RF@restore}
\def\ref@restore{\ref@setup \ref@restore}
\def\RF@restore{\ref@setup \RF@restore}

\outer\def\REFext{\@ext2\REF@ext}
\def\REF@ext{\@@ext\ref@name\refn@m}


\newtoks\R@names   \R@names={}
\newcount\RFn@m   \newcount\RF@high
\newcount\RFmax   \RFmax=50  
\def\RF@type{RF}
\let\RF@list=\empty
\newwrite\RF@write  \def\RF@file{\RF@file\RF@write0}
\let\RF@noc=N

\begingroup \let\storefile=\relax        
  \let\RFlist=\relax \let\RFfile=\relax  
  \let\RFext=\relax \let\RF=\relax       
  \gdef\@RF{%
    \glet\RFlist\undefined
    \if L\RF@lfe \glet\RFlist\empty \glet\@RF\@RFlist
      \gdef\RF@input{\RF@list}%
      \else \glet\@RFlist\undefined \fi
    \glet\RFfile\undefined
    \if F\RF@lfe \glet\RFfile\empty \glet\@RF\@RFfile
      \else \RF@setup
      \glet\@RFfile\undefined \glet\@RFcopy\undefined
      \glet\RF@store\undefined \glet\RF@copy\undefined
      \glet\RF@@input\undefined \fi
    \glet\RFext\undefined
    \if E\RF@lfe \gdef\RFext##1 {}\glet\@RF\@RFext
      \else \glet\@RFext\undefined \fi
    \glet\RF@setup\undefined \@RF}
  \gdef\RF@setup{\ifx \storefile\undefined \glet\file@store\undefined
    \glet\file@open\undefined \glet\file@close\undefined
    \glet\file@wlog\undefined \glet\file@free\undefined
    \glet\file@copy\undefined \glet\file@read\undefined \fi}
\endgroup

\outer\def\RFlist{\glet\RF@lfe L}
\outer\def\RFfile{\if B\store@blf \opt@err \else \glet\RF@lfe F\fi}
\outer\def\RFext#1 {\if B\store@blf \opt@err \else
  \glet\RF@lfe E\gdef\RF@input{\input#1 }\RF@input \fi}

\def\@RFdef#1{\gdef#1{\RF@def#1}}
\def\@RF@list#1{\toks@\expandafter{\RF@list\RF@#1}%
  \xdef\RF@list{\the\toks@ {\the\toks@store}}}
\def\@RFcopy#1{\RF@store{\noexpand#1}}
\def\RF@store{\let\@type\RF@type \if C\RF@noc \glet\RF@noc N%
  \RF@copy \fi \glet\RF@noc O\expandafter\file@store\RF@file}
\def\RF@input{\expandafter\RF@@input\RF@file}
\begingroup \let\RF=\relax  
  \gdef\RF@copy{{\let\\\RF \let\@RF\@RFcopy
    \expandafter\file@copy\RF@file}}
  \gdef\RF@@input#1#2#3{\if O\RF@noc \let\@type\RF@type
    \file@close#2#3\glet\RF@noc C\fi \let\\\RF \file@read#3}
\endgroup

\def\RF@def#1{\ref@@name\R@names#1\RF@def@ #1}
\def\RF@def@{\ref@setup \RF@def@}
\def\RF@def@num{\toks@store\expandafter{\expandafter\RF@find
  \expandafter{\the\refn@m}}\@@ref}
\def\RF@def@nam{\ifnum\refn@m=\@ne \refadd{\RF@print}\fi}
\def\RF@test#1{\z@}

\def\RF@print{\let\@RF\@RF@print \let\RF@def\RF@test
  \let\RF@first T\RF@input}
\def\@RF@print#1{\ifnum#1>\z@
  \if\RF@first T\let\RF@first F\setbox\z@\lastbox
  \else \form@t\f@rmat \noindent \strut \fi
  \hangindent\namrefindent \the\toks@store \fi}

\def\RF@find#1{\RFn@m#1\bgroup \let\RF@def\RF@test
  \if L\RF@lfe \else \ifnum\RFn@m<\RF@high \else \RF@search \fi \fi
  \let\RF@\RF@locate \RF@list \egroup}

\def\RF@search{\global\RF@high\RFn@m \global\advance\RF@high\RFmax
  \glet\RF@list\empty \let\@RF\@RFread \let\par\relax \RF@input}
\def\@RFread#1{\ifnum#1<\RFn@m \else \ifnum#1<\RF@high
  \@RF@list#1\fi \fi}
\def\RF@locate#1#2{\ifnum#1=\RFn@m #2\fi}

\def\@RFlist#1{\@RFdef#1\@RF@list#1}
\def\@RFfile#1{\@RFdef#1\@RFcopy#1}
\let\@RFext=\@RFdef

\outer\def\RF{\num@lett\RF@}
\def\RF@#1{\ref@unskip \read@store{\@RF#1}}

\outer\def\yearpage{\glet\yearpage@yp Y}
\outer\def\pageyear{\glet\yearpage@yp P}

\def\journal#1{{\journalstyle{#1}}\j@urnal{}}
\def\journalp#1{{\journalstyle{#1}}\j@urnal}
\def\journalf#1#2#3({{\journalstyle{#1}}\j@urnal{#3}#2(}

\def\j@urnal#1#2(#3)#4*{\unskip
  \ {\volumestyle{#1\ifx @#1@\else\ifx @#2@\else
  \kern.2em\fi \fi#2}}\unskip
  \ifx @#3@\else\ifx @#4@ (#3)\else\if Y\yearpage@yp\ (#3) #4\else
  , #4 (#3)\fi \fi \fi}

\def\Lett{Lett.\ }

\def\Phys{Phys.\ }

\def\PLB{\journalf{\Phys\Lett}B}


\newcount\tocn@m   \tocn@m=0
\newcount\auto@toc   \auto@toc=-1
\let\toc@saved\empty

\def\toc@type{toc}
\newbox\toc@box
\let\toc@list=\empty
\newwrite\toc@write  \def\toc@file{\toc@file\toc@write0}

\outer\def\tocpage{\glet\toc@page T}
\outer\def\notocpage{\glet\toc@page F}

\outer\def\tocnone{\gm@ne\auto@toc}
\outer\def\tocchap{\gz@\auto@toc}
\outer\def\tocsect{\global\auto@toc\@ne}

\def\toc#1{\read@store{\@toc{#1}}}
\def\tocadd{\@add3}
\def\@toc{\g@ne\tocn@m
  \expandafter\@@toc\csname toc@\romannumeral\tocn@m\endcsname}
\def\@@toc#1{\pagelabel#1%
  \toks@store\expandafter{\the\toks@store\toc@fill#1}\@store3}
\def\toc@fill{\rightskip4em\@plus1em\@minus1em\parfillskip-\rightskip
  \unskip\vadjust{}\leaders\hbox to1em{\hss.\hss}\hfil}

\outer\def\tocout{\@out3\toc@page\tocpref\tochead\emptyt@ks}
\outer\def\tockill{\@kill3}
\outer\def\restoretoc#1{\read@@store{\@store3{#1}}}

\newcount\footn@m   \footn@m=0
\def\foot@type{foot}
\newbox\foot@box
\let\foot@list=\empty
\newwrite\foot@write  \def\foot@file{\foot@file\foot@write0}

\outer\def\footsqb{\glet\foot@bp B\glet\thefootn@m\sqbfootn@m}
\outer\def\footpar{\glet\foot@bp P\glet\thefootn@m\parfootn@m}

\outer\def\footbot{\glet\foot@be B\glet\vfootnote\vfootn@te}
\outer\def\footend{\glet\foot@be E\glet\vfootnote\foot@store}

\outer\def\footpage{\glet\foot@page T}
\outer\def\nofootpage{\glet\foot@page F}

\def\sqbfootn@m{\lbrack \the\footn@m \rbrack}
\def\parfootn@m{\the\footn@m )}

\def\foot{\hfoot \vfootnote\footid}
\def\hfoot{\g@ne\footn@m \edef\n@xt{{$^{\thefootn@m}$}}%
  \expandafter\hfootnote\n@xt}
\def\footnote#1{\hfootnote{#1}\vfootnote\footid}
\def\hfootnote#1{\let\@sf\empty
  \ifhmode\unskip\edef\@sf{\spacefactor\the\spacefactor}\/\fi
  #1\@sf \gdef\footid{#1}}
\def\vfootn@te#1{\insert\footins\bgroup \foot@style
  \llap{#1}\after@arg\@foot}
\let\fo@t=\undefined
\let\f@@t=\undefined
\let\f@t=\undefined

\def\foot@style{\footstyle  
  \interlinepenalty\interfootnotelinepenalty
  \baselineskip\footnotebaselineskip
  \splittopskip\interfootnoteskip 
  \splitmaxdepth\dp\strutbox \floatingpenalty\@MM
  \leftskip.05\hsize \rightskip\z@skip
  \spaceskip\z@skip \xspaceskip\z@skip \noindent \footstrut}

\def\foot@store#1{\read@store{\@store4{#1}}}
\def\footadd{\@add4}

\outer\def\footout{\@out4\foot@page\footpref\foothead\emptyt@ks}
\outer\def\footkill{\@kill4}
\outer\def\restorefoot#1{\read@store{\@store4{#1}}}

\def\ignorefoot{\let\foot\eat \let\hfoot\eat  
  \def\footnote{\expandafter\eat\eat}\let\hfootnote\footnote}


\def\varitem{\afterassignment\v@ritem \setbox\z@\hbox}
\def\v@ritem{\hss \bgroup \aftergroup\v@@ritem}
\def\v@@ritem{\enskip \egroup \endgraf \noindent
  \hangindent\wd\z@ \box\z@ \ignorespaces}

\def\hvskip{\afterassignment\h@vskip \skip@}
\def\h@vskip{\unskip\nobreak \vadjust{\vskip\skip@}\lb \ignorespaces}

\def\parvskip{\bgroup \afterassignment\par@vskip \parskip}
\def\par@vskip{\parindent\hangindent \endgraf \indent \egroup
  \ignorespaces}

\def\item{\varitem to2.5em}
\def\sitem{\varitem to4.5em}
\def\ssitem{\varitem to6.5em}


\newcount\pointn@m   \pointn@m=0
\def\pointbegin{\gz@\pointn@m \point}
\def\point{\g@ne\pointn@m
  \xdef\the@label{\the\pointn@m}\item{\the@label.}}

\newcount\spointn@m   \spointn@m=96
\def\spointbegin{\global\spointn@m96 \spoint}
\def\spoint{\g@ne\spointn@m
  \xdef\the@label{\char\the\spointn@m}\sitem{(\the@label)}}

\newcount\sspointn@m   \sspointn@m=0
\def\sspointbegin{\gz@\sspointn@m \sspoint}
\def\sspoint{\g@ne\sspointn@m
  \xdef\the@label{\romannumeral\sspointn@m}\ssitem{\the@label)}}



\def\matc{\let\mat@lfil\hfil \let\mat@rfil\hfil}
\def\matl{\let\mat@lfil\relax \let\mat@rfil\hfil}
\def\matr{\let\mat@lfil\hfil \let\mat@rfil\relax}

\def\matrix#1{\null\,\vcenter{\normalbaselines\m@th
    \ialign{$\mat@lfil##\mat@rfil$&&\quad$\mat@lfil##\mat@rfil$\crcr
      \mathstrut\crcr\noalign{\kern-\baselineskip}%
      #1\crcr\mathstrut\crcr\noalign{\kern-\baselineskip}}}\,}

\def\bordermatrix#1{\begingroup \m@th
  \setbox\z@\vbox{%
    \def\cr{\crcr\noalign{\kern2\p@\glet\cr\endline}}%
    \ialign{$##\hfil$\kern2\p@\kern\p@renwd&\thinspace$\mat@lfil##%
      \mat@rfil$&&\quad$\mat@lfil##\mat@rfil$\crcr
      \omit\strut\hfil\crcr\noalign{\kern-\baselineskip}%
      #1\crcr\omit\strut\cr}}%
  \setbox\tw@\vbox{\unvcopy\z@\global\setbox\@ne\lastbox}%
  \setbox\tw@\hbox{\unhbox\@ne\unskip\global\setbox\@ne\lastbox}%
  \setbox\tw@\hbox{$\kern\wd\@ne\kern-\p@renwd\left(\kern-\wd\@ne
    \global\setbox\@ne\vbox{\box\@ne\kern2\p@}%
    \vcenter{\kern-\ht\@ne\unvbox\z@\kern-\baselineskip}\,\right)$}%
  \null\;\vbox{\kern\ht\@ne\box\tw@}\endgroup}


\mathchardef\smallsum=\dq1006
\mathchardef\smallprod=\dq1005

\def\b@mmode{\relax\ifmmode \expandafter\c@mmode \else $\fi}
\def\c@mmode#1\e@mmode{#1}
\def\e@mmode{$}
\def\defmmode#1#2{\def#1{\b@mmode#2\e@mmode}}

\defmmode\{{\lbrace}
\defmmode\}{\rbrace}

\defmmode\,{\mskip\thinmuskip}
\defmmode\>{\mskip\medmuskip}
\defmmode\;{\mskip\thickmuskip}

\defmmode{\Mit#1}{\mit#1}
\defmmode{\Cal#1}{\cal\uppercase\expandafter{#1}}

\def\dotii#1{{\mathop{#1}\limits^{\vbox to -1.4\p@{\kern-2\p@
   \hbox{\tenrm..}\vss}}}}
\def\dotiii#1{{\mathop{#1}\limits^{\vbox to -1.4\p@{\kern-2\p@
   \hbox{\tenrm...}\vss}}}}
\def\dotiv#1{{\mathop{#1}\limits^{\vbox to -1.4\p@{\kern-2\p@
   \hbox{\tenrm....}\vss}}}}

\let\barsymbol -
\mathchardef\tildesymbol=\dq0218
\def\hatsymbol{{\mathchoice{\null}{\null}{\,\,\hbox{\lower 10\p@\hbox
    {$\widehat{\null}$}}}{\,\hbox{\lower 20\p@\hbox
       {$\hat{\null}$}}}}}


\def\begin@stmt{\par\noindent\bpargroup\stmttitlestyle}
\def\adv@stmt#1#2#3#4{\begin@stmt
  \count@\ifx#2#3#1 \else\z@ \glet#2#3\fi \advance\count@\@ne
  \xdef#1{\the\count@}\edef\the@label{#4#1}}

\def\make@stmt{\ \the@label \make@@stmt}
{\catcode`\:\active
  \gdef\make@@stmt{\ \catcode`\:\active \let:\end@stmt}
}
\def\end@stmt{\catcode`\:\@ther \unskip :\stmtstyle
  \enskip \ignorespaces}

\def\defstmt#1#2#3{\expandafter\def@stmt \csname#1\endcsname
  {#2}{#1@stmt@}#3@}
\def\def@stmt#1#2#3#4#5@{\bgroup
  \toks@{\begin@stmt #2\make@@stmt}\toks@ii{#2\make@stmt}%
  \if#4n\xdef#1{\the\toks@}\else
    \xdef#1{\csname#3adv\endcsname \the\toks@ii}%
    \if#4=\edef\n@xt{\expandafter\noexpand
      \csname#5@stmt@adv\endcsname}\else
      \toks@{\empty}\toks@ii\toks@ \if#4c\toks@{\dot@pref}\fi
      \if#4s\toks@{\sect@pref}\if#5c\toks@ii{\sect@dot@pref}\fi \fi
      \if#5a\toks@ii\toks@ \fi
      \edef\n@xt{\noexpand\adv@stmt \csname#3num\endcsname
        \csname#3save\endcsname \the\toks@ \the\toks@ii}\fi
    \expandafter\glet\csname#3adv\endcsname\n@xt \fi
  \egroup}

\def\Prf{\par\noindent\bpargroup\prftitlestyle \case@language\prfhead
  \let\stmtstyle\prfstyle \make@@stmt}

\def\save@type{.texsave }  
\newread\test@read
\newwrite\save@write

\def\s@ve{\immediate\write\save@write}
\begingroup \catcode`\:=\active \catcode`\;=\active
  \outer\gdef\save#1 {{\let\,\space
    \immediate\openout\save@write#1\save@type
    \s@ve{;* definitions for :restore #1 \date\space- \thetime\space*}%
    \s@ve{:comment}%
    \s@ve{:\case@language{german\else english}}\save@page \save@chap
    \save@equ \save@fig \save@tab \save@ref \save@toc \save@foot
    \bgroup \def\n@xt##1.##2{\advance##2\@ne \s@ve{:start##1\the##2}}%
      \n@xt chap.\chapn@m \n@xt sect.\sectn@m
      \n@xt appendix.\appn@m \n@xt equ.\eqn@m
      \n@xt fig.\fign@m \n@xt tab.\tabn@m
      \n@xt ref.\refn@m \n@xt toc.\tocn@m
      \n@xt foot.\footn@m \egroup
    \s@ve{:\ifx\chap@@eq\sect@@eq app\else
      \ifnum\chapn@m=\z@ sect\else chap\fi \fi init}%
    \@save0\@save1\@save2\@save3\@save4%
    \if K\ref@kc \s@ve{:endcomment}\fi
    \save@restore ref \r@names  \save@restore RF \R@names
    \if C\ref@kc \s@ve{:endcomment}\fi
    \save@restore lbl \l@names  \save@restore eq \e@names
    \save@restore fig \f@names  \save@restore tab \t@names
    \s@ve{;*  end of definitions  *}\immediate\closeout\save@write
    }\wlog{* file #1\save@type saved *}}
  \gdef\save@restore#1 {\def\\{\s@ve{:#1restore}%
      \let\\\save@@restore \\}\the}
  \gdef\save@@restore#1{\toks@\expandafter{#1}%
    \s@ve{:dorestore\string#1{\the\toks@}}}
  \gdef\save@page{\s@ve{:\@opt\page@tbn Ttop Bbot Nno *pagenum%
      :\@opt\head@lrac Llef Rrigh Aal Ccen *thead%
      :\@opt\foot@lrac Llef Rrigh Aal Ccen *tfoot%
      :page\@opt\page@ac Aall Cchap *%
      :\@opt\ori@pl Pportrait Llandscape *}%
    \s@ve{:startpage\@opt\page@ac C\the\pageno@pref. *\the\pageno}}
  \gdef\save@chap{\s@ve{:\@opt\chap@page Fno *chappage%
      :\@opt\chap@yn Nno *chapters%
      :\@opt\chap@ar Aarabic Rroman *chapnum}}
  \gdef\save@equ{\s@ve{:equ\@opt\eq@acs Aall Cchap Ssect *%
      :equ\@opt\eq@lrn Lleft Rright Nnone *%
      :equ\@opt\eq@fs Ffull Sshort *}}
  \gdef\save@fig{\s@ve{:\@opt\fig@page Fno *figpage%
      :fig\@opt\fig@acs Aall Cchap Ssect *}}
  \gdef\save@tab{\s@ve{:\@opt\tab@page Fno *tabpage%
      :tab\@opt\tab@acs Aall Cchap Ssect *}}
  \gdef\save@ref{\s@ve{:\@opt\ref@page Fno *refpage%
      :ref\@opt\ref@kc Kkeep Cclear *%
      :ref\@opt\ref@sbn Ssup Bsqb Nnam *%
      :\@opt\yearpage@yp Yyearpage Ppageyear *}}
  \gdef\save@toc{\s@ve{:\@opt\toc@page Fno *tocpage%
      :toc\ifcase\save@@toc \else none\fi}}
  \gdef\save@foot{\s@ve{:\@opt\foot@page Fno *footpage%
      :foot\@opt\foot@be Bbot Eend *%
      :foot\@opt\foot@bp Bsqb Ppar *}}
\endgroup
\def\@opt#1#2#3 #4{\if #1#2#3\fi \if #4*\else
  \expandafter\@opt\expandafter#1\expandafter#4\fi}
\def\save@@toc{\auto@toc chap\or sect}

\newif\ifcr@ss
\begingroup \let\comment=\relax
  \outer\gdef\restore{\@kill0\@kill1\@kill2\@kill3\@kill4%
    {\def\\##1{\glet##1\undefined}%
      \def\n@xt##1{\the##1\global##1\emptyt@ks}%
      \n@xt\l@names \n@xt\e@names \n@xt\f@names \n@xt\t@names
      \n@xt\r@names \let\\\@RFdef \n@xt\R@names}
    \bgroup \let\comment\relax \let\d@rest@re\dorest@re \@restore}
\endgroup
\def\crossrestore#1 {\bgroup \openin\test@read#1\save@type
  \ifeof\test@read \let\n@xt\egroup
    \message{* file #1\save@type missing *}%
  \else \closein\test@read
    \def\n@xt{\@restore#1 }\let\d@rest@re\docr@ss \fi \n@xt}
\def\@restore#1 {\input #1\save@type \egroup}
\def\all@restore{\glet\name@list}
\outer\def\dorestore{\bgroup \catcode`\@\l@tter \num@lett\d@restore}
\def\d@restore#1#2{\egroup \toks@{#2}\d@rest@re#1}
\def\dorest@re#1{\def@name\name@list#1{\the\toks@}}
\def\docr@ss#1{\ifx#1\undefined \cr@sstrue
    \else \expandafter\testcr@ss#1\cr@ss\@@ \fi
  \ifcr@ss \expandafter\testcr@ss\the\toks@\cr@ss\@@ \else \cr@sstrue \fi
  \ifcr@ss\else \expandafter\dorest@re\expandafter#1\fi}
\def\testcr@ss#1\cr@ss#2\@@{\ifx @#2@\toks@{\cr@ss#1}\cr@ssfalse
  \else \cr@sstrue \fi}
\let\cr@ss=\empty

\def\def@name#1#2{\let\name@list#1\add@name#2\xdef#2}
\def\del@name#1{\bgroup
  \def\n@xt##1\\#1##2\\#1##3\@@##4{\global##4{##1##2}}%
  \def\del@@name##1{\expandafter\n@xt\the##1\\#1\\#1\@@##1}%
  \del@@name\l@names \del@@name\e@names \del@@name\f@names
  \del@@name\t@names \del@@name\r@names \del@@name\R@names \egroup}
\def\add@name#1{\del@name#1%
  {\global\name@list\expandafter{\the\name@list\\#1}}}

\def\kill{\num@lett\@k@ll}
\def\@k@ll#1{\def\k@ll##1{\ifx##1\k@ll \let\k@ll\relax \else
    \del@name##1\glet##1\undefined \fi \k@ll}\k@ll#1\k@ll}


\outer\def\startchap{\st@rt\chapn@m}
\outer\def\startsect{\st@rt\sectn@m}
\outer\def\startappendix{\st@rt\appn@m}
\outer\def\startequ{\st@rt\eqn@m}
\outer\def\startfig{\st@rt\fign@m}
\outer\def\starttab{\st@rt\tabn@m}
\outer\def\startref{\st@rt\refn@m}
\outer\def\starttoc{\st@rt\tocn@m}
\outer\def\startfoot{\st@rt\footn@m}

\def\st@rt#1{\gm@ne#1 \global\advance#1}



\hoffset@corr@p=86mm   \voffset@corr@p=98mm      
\hoffset@corrm@p=-5mm   \voffset@corrm@p=4mm      
\hoffset@corr@l=134mm   \voffset@corr@l=70mm     
\hoffset@corrm@l=-5mm   \voffset@corrm@l=-4mm     


\def\wlog#1{} 
\catcode`\@=11


\def\({\relax\ifmmode[\else$[$\nobreak\hskip.3em\fi}
\def\){\relax\ifmmode]\else\nobreak\hskip.2em$]$\fi}

\def\gappr{\mathpalette\under@rel{>\approx}}
\def\lappr{\mathpalette\under@rel{<\approx}}
\def\gsim{\mathpalette\under@rel{>\sim}}
\def\lsim{\mathpalette\under@rel{<\sim}}
\def\under@rel#1#2{\under@@rel#1#2}

\def\under@@rel#1#2#3{\mathrel{\mathop{#1#2}\limits_{#1#3}}}

\def\under@@rel#1#2#3{\mathrel{\vcenter{\hbox{$%
  \lower3.8pt\hbox{$#1#2$}\atop{\raise1.8pt\hbox{$#1#3$}}%
  $}}}}

\def\mppae@text{{Max-Planck-Institut f\"ur Physik und Astrophysik}}
\def\mppwh@text{{Werner-Heisenberg-Institut f\"ur Physik}}

\def\mppaddresstext{Postfach 40 12 12, D-8000 M\"unchen 40\else
  P.O.Box 40 12 12, Munich (Fed.^^>Rep.^^>Germany)}

\def\mppaddress{\address{\mppae@text \nl -- \mppwh@text\space --\nl
  \case@language\mppaddresstext}}


\font\fourteenssb=cmssdc10 scaled \magstep2 
\font\seventeenssb=cmssdc10 scaled \magstep3 

\def\letter#1#2{\b@lett@r{26}%
  \centerline{\seventeenssb \uppercase\mppae@text}%
  \centerline{\fourteenssb \uppercase\mppwh@text}%
  \centerline{\strut#1}\vskip.5cm%
  \e@lett@r{\hss\vtop to5cm{\hsize55mm%
    \lftline{\strut}\eightrm  \setbaselineskip=12pt \vfil
    \lftline{F\"OHRINGER RING 6}\lftline{\tenrm D-8000 M\"UNCHEN 40}%
    \lftline{\case@language{TELEFON\else PHONE}: (089) 32 308
      \if!#2!\else - #2 \case@language{oder\else or} \fi-1}%
    \lftline{TELEGRAMM:}\lftline{PHYSASTROPLANCK M\"UNCHEN}%
    \lftline{TELEX: 52 15 61 9 mppa d}%
    \lftline{POSTFACH 40 12 12}
    \ifx\EARN\undefined\else\vskip5\p@\lftline{EARN/BITNET: \EARN
      @DM0MPI11}\fi \vfil}}}

\def\b@lett@r#1{\endpage \begingroup \doublespace \vglue-#1mm}

\def\e@lett@r#1#2{\skippagenum T\skipheadline T\skipfootline T%
  \line{\vtop to47mm{\lftline{\llap{\vbox to\z@{\vskip171\p@
      \hrule\@width7\p@\vss}\hskip57\p@}\strut}\vskip2mm\vfil
    \addressspacing \dimen@\baselineskip \dimen@ii-2.79ex%
    \advance\dimen@ii\dimen@ \baselineskip\dimen@\@minus\dimen@ii
    \let\nl\cr\use@nl \halign{##\hfil\crcr#2\crcr}\vfil}#1}%
  \vskip1cm\rtline{\thedate}\vskip1cm\@plus1cm\@minus.5cm\endgroup}

\let\addressspacing=\empty


\def\myname{Dr.\ Xxxx Xxxxxxxxxx\nl Physiker}
\def\myaddress{Xxxxxxx Stra\ss e  ??\nl
    \llap{D--8000\quad}M\"unchen ??\nl
    Tel:\ (089) \vtop{\hbox{?? ?? ?? (privat)}%
                      \hbox{3 18 93-??? (B\"uro)}}}
\def\myletter{\b@lett@r{26}\line{\let\nl\cr \use@nl \caps
  \vtop to25mm{\halign{\strut##\hfil\crcr\myname\crcr}\vfil}\hfil
  \vtop to25mm{\tenpoint
    \halign{\strut##\hfil\crcr\myaddress\crcr}\vfil}}%
  \e@lett@r\empty}


\def\firstpageoutput{\physoutput
  \global\output{\setbox\z@\box@cclv \deadcycles\z@}}


\def\veq{\afterassignment\v@eq \dimen@}
\def\v@eq{$$\vcenter to\dimen@{}$$}

\def\veqn{\afterassignment\v@eqn \dimen@}
\def\v@eqn{$$\vcenter to\dimen@{}\eqn$$}

\def\heq{\afterassignment\h@eq \dimen@}
\def\h@eq{$\hbox to\dimen@{}$ }

\def\wlog{\immediate\write\m@ne} 
\catcode`\@=12 
\def\wlog#1{} 
\catcode`\@=11

\def\wlog{\immediate\write\m@ne} 
\catcode`\@=12 


%
%
\def\wlog#1{} 
\catcode`\@=11


\font\CERNfont=cernlogo        
\font\headlinefont=cmssdc10 scaled \magstep1
\font\headeraddressfont=cmss10

\def\CERNlogo{\hbox{\CERNfont C}}            
\def\smallCERNlogo{\hbox{\CERNfont c}}       


\def\header{\setbox0=\vtop{\logo}
            \setbox1=\vtop{\vglue -16mm
                           \ialign{\hfil##\hfil\cr
                           \headlinefont
                  EUROPEAN ORGANIZATION FOR NUCLEAR RESEARCH\cr}}
            \line{\hglue -0.7cm\box0\hskip 16mm\box1\hss}\hfil}

\def\letterhead{\setbox0=\vtop{\logo}

\setbox1=\vtop{\baselineskip=13pt\lineskip=1pt\lineskiplimit=0pt
                               \vglue -18mm
                               \ialign{\hfil##\hfil\cr
                               \headlinefont
               ORGANISATION EUROP\'EENNE POUR LA RECHERCHE NUCL\'EAIRE\cr
                               \headlinefont
                  EUROPEAN ORGANIZATION FOR NUCLEAR RESEARCH\cr
                \noalign{\smallskip}
                               \headeraddressfont
               Laboratoire Europ\'een pour la Physique des Particules\cr
                \noalign{\vskip -0.5mm}
                               \headeraddressfont
                    European Laboratory for Particle Physics\cr}}
                \line{\hglue -0.7cm\box0\hskip 1cm\box1\hss}\hfil}

\def\CERNheader{\let\logo=\CERNlogo\header}
\def\CERNletterhead{\let\logo=\CERNlogo\letterhead}

\def\smallCERNheader{\let\logo=\smallCERNlogo\header}
\def\smallCERNletterhead{\let\logo=\smallCERNlogo\letterhead}


\def\letter#1#2{\b@lett@r{23}%
  \line{\hglue -5mm\CERNletterhead\hss}\vskip 1mm%
  \centerline{\strut#1}\vskip.5cm%
  \e@lett@r{\hss\vtop to5cm{\hsize55mm%
    \lftline{\strut}\eightrm  \setbaselineskip=12pt \vfil
    \lftline{POSTAL ADDRESS:}%
    \lftline{\ TH Division, CERN}\lftline{\ CH-1211 Geneva 23}%
    \lftline{PHONE: (022) 767 #2}%
    \lftline{TELEFAX: (022) 782 39 14}%
    \lftline{TELEGRAMME:}\lftline{CERNLAB-GEN\`EVE}%
    \lftline{TELEX: 2 36 98 CH}%
    \ifx\EARN\undefined\else\vskip5\p@\lftline{EARN/BITNET: \EARN
      @CERNVM}\fi \vfil}}}

\def\e@lett@r#1#2{\skippagenum T\skipheadline T\skipfootline T%
  \line{\vtop to47mm{\lftline{\llap{\vbox to\z@{\vskip171\p@
      \hrule\@width7\p@\vss}\hskip57\p@}\strut}\vskip2mm\vfil
    \addressspacing \dimen@\baselineskip \dimen@ii-2.79ex%
    \advance\dimen@ii\dimen@ \baselineskip\dimen@\@minus\dimen@ii
    \vbox{\overfullrule 0pt\hsize=7cm\parskip=0pt\parindent=0pt
          \def\nl{\par\hangindent=1.5em\hangafter=1 }#2}\vfil}#1}%
    \vskip1cm\rtline{\thedate}\vskip1cm\@plus1cm\@minus.5cm\endgroup}


\def\preprintheader{\vglue -23mm\line{\hglue -5mm\CERNheader\hss}%
                    \vglue -8mm}
\def\pubnum#1{\topright{CERN-TH.#1}\def\thepubnum{#1}}
\def\pubdate#1{\def\thepubdate{#1}}
\def\CERNaddress{\address{Theory Division, CERN\nl
                          CH-1211 Geneva 23, Switzerland}}
\def\endtitlepage{\footline={\vtop{\null\vfil
                     \hbox{\ifx\thepubnum\undefined
                           \else CERN-TH.\thepubnum\fi}
                     \hbox{\ifx\thepubdate\undefined
                             \ifcase\month\or
                                 January\or February\or March\or
                                 April\or May\or June\or
                                 July\or August\or September\or
                                 October\or November\or December\fi,
                           \space\number\year
                           \else\thepubdate\fi}\vss}
                           \hfil}\endpage
\footline={\footlinestyle \head@foot\skip@foot\foot@lrac
  \lfoottext\cfoottext\rfoottext\footb@x}\pageno=1}


\def\PLB#1(#2)#3*{\ifnum #1<171
                      \journalf{\Phys\Lett}B#1(#2)#3*%
                  \else
                      \journalp{\Phys\Lett}B#1(#2)#3*%
                  \fi}
\def\ZPC{\journalp{Z.\ \Phys}C}               

%

%
\def\thedate{\number\day\case@language{.\ \else\ }%
             \themonth\case@language{\else,} \number\year}


\def\today{\number\day\space\ifcase\month\or
 January\or February\or March\or April\or May\or June\or
 July\or August\or September\or October\or November\or December\fi,
 \number\year}

\long\def\blankpage{
     \skippagenum=T   \skipheadline=T   \skipfootline=T
     \null\vfill\eject
     \advance\pageno -1
     \skippagenum=F   \skipheadline=F   \skipfootline=F}

\def\wlog{\immediate\write\m@ne} 
\catcode`\@=12 
\catcode`\@=12 

\message{and default options.}



\hbadness=2000  

\newsect        

\english        

\botpagenum     
\centhead       
\centfoot       
\pageall        
\portrait       
\titlepage      
\nochappage     
\arabicchapnum  
\chapters       
\equchap        
\equshort       
\equright       
\storelist      
\figall         
\figpage        
\nographics     
\taball         
\tabpage        
\refsup         
\refpage        
\refkeep        
\RFlist         
\yearpage       
\tocpage        
\tocnone        
\footsqb        
\footbot        
\footpage       
\matc           

\wlog{summary of allocations:}
\wlog{last count=\number\count10 }
\wlog{last dimen=\number\count11 }
\wlog{last skip=\number\count12 }
\wlog{last muskip=\number\count13 }
\wlog{last box=\number\count14 }
\wlog{last toks=\number\count15 }
\wlog{last read=\number\count16 }
\wlog{last write=\number\count17 }
\wlog{last fam=\number\count18 }
\wlog{last insert=\number\count19 }

%
%
%
%
%
%
%
\equfull
\refsqb
\overfullrule 0pt
\preprintheader
\pubnum{6449/92}
\pubdate{April, 1992}
\title{Two--Body Decays of $B_s$ Mesons }
\author{J. Bijnens}
\autcon{F. Hoogeveen}
\CERNaddress
\abstract{
We have calculated the decay rates of the $B_s$ meson in a number of
exclusive two--body decay channels using
 the Bauer--Stech--Wirbel model for
current matrix elements.
The influence of the free parameters of the model on the predictions is
studied.
The total branching ratio of the $B_s$ into
 final states which only contain
stable charged particles is found to be about $10^{-3}$. }
\endtitlepage
\RF\BSW{M.Bauer, B.Stech, M.Wirbel, \ZPC34(1987)103*}
\RF\WSB{M.Wirbel, B.Stech, M.Bauer, \ZPC29(1985)637*}
\RF\ARGUS{H.Albrecht et al. [ARGUS collab.], \ZPC 48(1990)543*}
\RF\CLEO{ D.Borteletto et al. [CLEO collab.],
 Inclusive and exclusive decays of $B$--mesons to final states
 including charm and charmonium mesons, CLNS-91-1102}
\RF\CPVIOL{CP violation, ed. C.Jarlskog, World Scientific,
           Singapore, 1989}
\RF\PDG{Part. Data Group, \PLB 239(1990)1*}
\RF\FIT{M.Neubert et al., Exclusive Weak decays of B--Mesons,
        HD-THEP-91-28}
\def\Dsxp{$D_s^{*+}$}
\def\Etap{$\eta^\prime$}
\def\Kxp{$K^{*+}$}
\def\Kxm{$K^{*-}$}
\def\Dsp{$D_s^+$}
\def\Etaa{$\eta$}
\def\Dsxm{$D_s^{*-}$}
\def\Phii{$\phi$}
\def\Kz{$K^0$}
\def\Aim{$a_1^-$}
\def\Kxz{$K^{*0}$}
\def\Kp{$K^+$}
\def\Rhom{$\rho^-$}
\def\Dsm{$D_s^-$}
\def\Pim{$\pi^-$}

\def\Jpsi{$J/\psi$}
\def\Dmx{$D^{*-}$}
\def\Kmx{$K^{*-}$}
\def\Dm{$D^-$}

\def\Km{$K^-$}
\def\Dzx{$D^{*0}$}
\def\Dz{$D^0$}
\def\Etac{$\eta_c$}
\def\Aiz{$a_1^0$}
\def\Rhoz{$\rho^0$}
\def\Dzbar{$\overline{D^0}$}
\def\Omeg{$\omega$}
\def\Dzxbar{$\overline{D^{*0}}$}
\def\Dm{$D^-$}
\def\Piz{$\pi^0$}
\def\Rhoz{$\rho^0$}

\chap{ Introduction.}
It does not have to be emphasized that the study of $B$--mesons is
worthwhile.
The observation of mixing and the prospect of observing  $CP$--violation
in the \hbox to 1 cm{}  \hbox{$B$--system}
 are exciting and may give deep insight into the inner
mechanism of the standard model.
Of the pseudoscalar $B$--mesons only the $B^\pm$ and the $B^0$ have been
seen in fully reconstructed events \quref{\ARGUS,\CLEO}.
So far the $B_s$--meson has remained elusive.
Although it would be highly surprising if this state didn't exist, it is
very important to actually find it.
The $B_s$ meson is expected to mix strongly with its anti particle and
many proposed schemes\quref\CPVIOL\ %
for detecting $CP$--violation in the $B$--system use this particle as
initial state.

In this paper we calculate a
number of branching ratios of the $B_s$--meson
into two--body channels.
To do so we employ the model of
Bauer, Stech and Wirbel\quref{\BSW,\WSB} which these authors
have
used to calculate the two body
decays of $D^\pm,D^0,B^\pm$ and $B^0$ mesons.

The organization of this paper is as follows.
In section 2 we briefly review the BSW model and discuss the
input parameters of the model.
In section 3 we give the results of our calculation together with a
discussion of the sensitivity of our results with respect to variations
in the input parameters.
Finally in section 4 we discuss goldplated decay chains where the
decay products at the end of the decay chain
 are all charged pions, kaons,
muons or electrons.
These can be used for the direct detection of the $B_s$.

\chap{ The Bauer--Stech--Wirbel model.}
With the current state of the art it is impossible to calculate matrix
element for processes involving strongly interacting particles from
first principles using the standard model.
Therefore one has to resort to models.
The model used here is a model by Bauer, Stech and Wirbel
\quref{\BSW,\WSB}.
This model can be used to calculate the decay of a spin $0$ meson in
two mesons of spin $0$ or $1$, as well as its semileptonic decay.
The two main ingredients are factorization and an {\it Ansatz\/} for the
current matrix elements.

The short distance effective action is given by
$$\eqalign{
 {\cal L}_{eff}=&
 {{G_F}\over \sqrt{2}}
   \big\{ c_1(\mu)({\bar u}d^\prime)({\bar s}^\prime c)
        + c_2(\mu)({\bar s}^\prime d^\prime)({\bar u} c)
         \big\} \cr
& + ((\bar{u}d^\prime)\to(\bar{c}s^\prime)) + \cdots  \cr}
                                               \EQN\ShortDistance  $$
where the primes on the quark fields denote the interaction eigenstates.
The factorization assumption amounts to replacing the short distance
effective action
by the following product of meson currents
$$ {\cal L}_{eff}= {{G_F}\over \sqrt{2}}
  : \big\{ a_1({\bar u}d^\prime)_H({\bar s}^\prime c)_H
        + a_2({\bar s}^\prime d^\prime)_H({\bar u} c)_H
         +\cdots
        \big\}:
                                              \EQN\LongDistance $$
where the index $H$ means that the quarks denoted inside the bracket are
to be contracted with one and the same meson.
In this way the problem of calculating decay matrix elements reduces to
the calculation of current matrix elements between a one meson state
and the vacuum or between two one meson states.
The former matrix elements are simply given by the corresponding
decay constants:
$$\langle 0 \vert J_\mu \vert A\,0^\pm \rangle = i f_A p_\mu \qquad
  \langle 0 \vert J_\mu \vert A\,1^\pm \rangle = \epsilon_\mu m_A F_A
                                                       \EQN\DecayConst $$
for a spin zero and a spin one particle respectively.
The latter kind of matrix element can be expressed in terms of
Lorentz--scalar formfactors:
$$\eqalign{
  \big\langle B \,0^\pm \vert J_\mu \vert A \,0^\pm \big\rangle=&
  \left(p_A+p_B-{{m_A^2-m_B^2}\over{q^2}}q\right)_\mu\, F_1(q^2) \cr
&+{{m_A^2-m_B^2}\over {q^2}}\,q_\mu \,F_0(q^2) \cr
  \big\langle B\, 1^\pm \vert J_\mu \vert A \,0^\pm \big\rangle &=
{2\over{m_A+m_B}}\epsilon_{\mu\nu\rho\sigma}\,\epsilon^{*\nu}
        p_A^\rho p_B^\sigma\, V(q^2) \cr
 &+i(m_A+m_B) \left(\epsilon^*_\mu-{{\epsilon^*\cdot q}\over{q^2}}
            q_\mu\right)\, A_1(q^2)\cr
  &-i\epsilon^*\cdot q
  \left({{p_{A\mu}-p_{B\mu}}\over{m_A+m_B}}-{{m_A-m_B}\over{q^2}}
  q_\mu\right)\,
      A_2(q^2) \cr
  &+i{{\epsilon^*\cdot q}\over{q^2}}2m_B\,q_\mu\, A_0(q^2)   }
                                                  \EQN\FormFactorDef $$
where $q=p_A-p_B$, and $\epsilon^\nu$ is the polarization vector of
the spin one meson.
The $q^2$ behaviour is assumed to be dominated
by the nearest pole which has
the correct quantum numbers.
At this point the third  ingredient
of the BSW model makes its appearance.
These authors assume that using the wavefunctions
of a relativistic harmonic
oscillator model provides a reasonable approximation for the wavefunction
of a physical meson.
This amounts to taking the following expressions
for the formfactors at zero momentum transfer :
$$\eqalign{
F_0(0)&=A_0(0)={f({{a_A+a_B}\over 2},{{b_A+b_B}\over 2} )\over
          \sqrt{f(a_A,b_A)f(a_B,b_B)}} \cr
A_1(0)&={{m_{A1}+m_{B1}}\over{m_A+m_B}}\cdot
        {{2\omega_A\omega_B}\over{\omega_A^2+\omega_B^2}}\cdot
               {g({{a_A+a_B}\over 2},{{b_A+b_B}\over 2} )\over
          \sqrt{f(a_A,b_A)f(a_B,b_B)}} \cr
V(0)&={{m_{A1}-m_{B1}}\over{m_A-m_B}}\cdot
        {{2\omega_A\omega_B}\over{\omega_A^2+\omega_B^2}}\cdot
               {g({{a_A+a_B}\over 2},{{b_A+b_B}\over 2} )\over
          \sqrt{f(a_A,b_A)f(a_B,b_B)}} \cr
A_2(0)&={{m_A+m_B}\over{m_A-m_B}}A_1(0)-{{2m_B}\over{m_A-m_B}}A_0(0).}
      \EQN\hi  $$
In this expression $m_{A,B}$ are the masses of the mesons
in the matrix element,
$m_{A1,B1}$ are  the masses of the nonspectator quarks, and $m_2$ is the
spectator quark mass.
The numbers $a_A$,$b_A$ are given by
$$ a_A={{m_A^2}\over{\omega_A^2}} \quad\quad
    b_A=-\,{{m_A^2+m_{A1}^2-m_2^2}\over {\omega_A^2}} \EQN\ab   $$
with similar expressions for $a_B$,$b_B$.
The functions $f$ and $g$, given by
$$\eqalign{
f(a,b)&=\int^1_0\!dx\,x(1-x)e^{-ax^2-bx} \cr
g(a,b)&=\int^1_0\!dx\,(1-x)e^{-ax^2-bx}} \EQN\fg $$
can be readily expressed in terms of the error function.
The parameters $\omega_A$ and $\omega_B$
define the spatial extensions
of the two mesons sandwiching the currents. (Small $\omega$ means a large
meson and vice versa).

Final state interactions play an important r\^ole in the analogous decays
of the $D$--meson, but for the $B_s$ meson they are expected to be small
due to the large mass of the $b$--quark.
In the following we have neglected all final state interactions.
Also neglected are the effects of weak annihilation contributions,
 penguin  diagrams and loop contributions.

Short distance QCD radiative corrections are in principle incorporated
in the coefficients $a_1,a_2$, but rather than calculating these numbers
from
first principles we are using the values which have been obtained
from a comprehensive fit to nonstrange $B$--meson decays\quref\FIT
$$  a_1 = 1.11   \qquad a_2 = 0.21      .\EQN\aNumbers $$

The values of the decay constants used can be found in table 2.1, whereas
the polemasses are listed in table 2.2 .
Many of these pole masses correspond to unobserved states.
In principle the
same values are being used as those of BSW,
unless more precise information
has become available in the mean time.

As constituent quark masses we assume 350 MeV for the up and down
quarks, 550 MeV for the
strange quarks and 1.7 GeV for the charm quark.
The b--quark mass was taken to be 4.9 GeV.

The values of the Kobayashi--Maskawa matrix elements were taken from
reference \quref\PDG, for $V_{ud},V_{us},V_{cd}$ and $V_{cs}$, whereas
$V_{cb}$ and $V_{ub}$ were taken to be $0.05$ and $0.005$ respectively.

Finally for the $B_s$ mass itself we adopted 5.4 GeV as a standard
value.

\chap{Results}
Going through the formalism
 described above and assuming that the lifetime
of the $B_s$--meson is the same as the average lifetime of the nonstrange
$B$--mesons $11.8\ 10^{-13}\rm \ sec$
we have calculated the branching ratios
for the $B_s$ into a number of two meson decay channels.
The results of our calculation can be found in table 3.1
There we list the branching ratio for
the decay, the KM matrix elements which
appear in the amplitude.
All two--body decay channels in the list
add up to a branching ratio of 11.4\%.
The branching ratio simply scales quadratically
with the KM matrix elements.
The decays into a pair of charged particles scale as $a_1^2$ and the
decays into a pair of neutral particles scale like $a_2^2$.
In this calculation all effects of mixing and/or CP violation are ignored.
If the $B_s$--meson is mixing strongly, as it is expected to do, then the
branching ratios for a mass eigenstate to
decay in a particular channel is just
the average of the branching ratios of
this channel and its charge conjugate.

The mass of the $B_s$--meson is presently not known.
For the numerical estimates in table 1 a
value of $5.4\rm\ GeV$ has been used.
Increasing this value to $5.5\rm\ GeV$ leads to an increase of all
branching  ratios considered.
The largest increase is by 29\% and occurs for the channel \Etap\Jpsi ,
whereas the smallest increase is 17\% which occurs in the channel
 \Etaa\Piz .
The changes in the decay channels with branching ratios over one promille
are more or less uniformly distributed between +27\% and +18\%.

For every particle the parameter $\omega$ is a free parameter
 defining the
spatial extension of the meson.
A value of $400\rm\ MeV$ gives a reasonable
description of $D$ and non--strange
$B$--meson decays, and has been adopted here for all particles.
Increasing the $\omega$ of the $B_s$--meson to $500\rm\ MeV$
 while keeping
the rest fixed has drastic consequences for some decay channels.
The branching ratio of the \Phii\Jpsi\ channel increases by 39\%.
The other important decay channels increase by between 20\% and 30\%.

The $q^2$ dependence of the formfactors is modeled using the assumption
that the nearest resonance with the correct quantum numbers dominates.
Many of the resonances which are needed are themselves not yet detected
however and consequently their mass is unknown.
The values from table 2.2 are used to produce the numbers in table 3.1.
To see the effect of this uncertainty,
we increased all pole masses by 10\%.
The resulting change in the branching
ratios was very small, most important
channels decreased by a few percent.
The \Phii\Jpsi\ channel was among a few
exceptions and decreased by 13\%.

The overall error of any calculation is very hard to estimate.
{}From the uncertainties in the input
parameters discussed above we know that
the inaccuracy can be at least as large as a factor of two for the decay
$B_s \to\phi J/\psi$ and as large as
 $1.5$ for the important decay channels
involving $D$--mesons.
The relative importance of the different
decay channels is however much less
influenced by the uncertainty in the input parameters.
The order of importance of the largest channels is not changed, even by
rather drastic changes in input parameters.

\chap{Goldplated decay chains.}
Decay chains in which the stable particles at the end of the chain are all
charged are especially suited for reconstructions of the $B_s$ meson.
So far not a single fully reconstructed $B_s$ has been seen.
The numerical values below have been taken from \quref\PDG.

First consider the decays of the $D_s^\pm$--meson.
In $2.7 \%$ of the time it decays in
$\phi\pi^\pm$ and $1.3\%$ of its decays
produce $\phi\pi^\pm\pi^\pm\pi^\mp$.
Of these roughly half are followed by the decay $\phi\to K^+K^-$.
The decay channel $K_SK^+$ followed by the decay of the $K_S$ into two
charged pions has a branching ratio of about 0.9\%.
Furthermore the decay channels $\overline{K^{*0}}K^+$
and $K^{*+}K_S$ have a probability of
$2.6\%$ and $1.6\%$ respectively.
The charged $K^*$ decays into a charged
pion and $K_S$ one third of
the time,
whereas the $K^{*0}$ decays into $K^+\pi^-$
 two thirds of the time.
Finally the direct decay $\pi^\pm\pi^\pm\pi^\mp$ has a branching ratio
of $1.2$\%.
The decay mode $K^+K^-\pi^+$ adds another 0.67 \%.
Taking everything together we notice that
the $D_s^\pm$ decays $6.7 \%$ of the
time into a final state in which all the particles are charged.

The $a_1^\pm$ meson decays with a branching ratio of fifty percent to
$\pi^\pm\pi^\pm\pi^\mp$  through  $\rho^0\pi^\pm$.

The \Jpsi\ decays into $e^+e^-$ or $\mu^+\mu^-$ with a branching ratio of
13.8\%.

Another interesting possibility is the $\psi^\prime$ which decays either
in charged leptons or into \Jpsi$\,\pi^+\pi^-$
followed by \Jpsi\ decaying into $e^+e^-$ or $\mu^+\mu^-$.
These channels contribute 6.1 \% to the all charged branching ratio
of the $\psi^\prime$.
The branching ratio $B_s \to \phi\psi^\prime$ has been calculated
using the same methods as in section 3.

All together we find
the branching ratios for goldplated decays
as they are displayed in table 4.1 .

Adding everything together we note that roughly one promille of the $B_s$
decays goes into a goldplated decay chain.
Furthermore to all these decay channels one may add $\pi^+\pi^-$ and or
$K^+K^-$ pairs ad libitum (or at least upto the kinematical limit).
This may bring a substantial improvement as a comparison with the
$D$--system shows.
$$
{{\Gamma(D^+\to\overline{K^0}\pi^+\pi^+\pi^-)
}\over{\Gamma(D^+\to\overline{K^0}\pi^+)}}=2.5
\quad\quad
{{\Gamma(D^0\to K^-\pi^+\pi^+\pi^-)}
\over{\Gamma(D^0\to K^-\pi^+)}}=2.1\ . \EQN\dratio
   $$.
\ack
The authors like to thank A.Pich and P.Kluit for a number of pleasant
discussions.
\refout
\break\vfil
\line{{\bf Table 2.1 :} The decay constants in MeV. \hfil }
\vskip 3 mm
\vbox{\offinterlineskip\hrule
\halign{
  \vrule\strut\quad # \hfil&\quad # \hfil\vrule&\quad # \hfil &\quad
      # \hfil\vrule \cr
  \noalign{\hrule}
  $\pi^\pm$           & 131.7 & $\rho,a_1,K^*,D^*,D_s^*$ & 221 \cr
  $K,D,D_s,\eta_c$    & 160.6 & $\omega$                 & 156 \cr
  $\eta\ (uu) $       &  94   & $\phi$                   & 233 \cr
  $\eta^\prime\ (uu)$ &  65   & \Jpsi                    & 382 \cr
  \noalign{\hrule} }}
\vskip 1 cm
\line{{\bf Table 2.2 :} The pole masses in MeV. \hfil}
\vskip 3 mm
\vbox{\offinterlineskip\hrule
 \halign{\vrule\strut\quad # \quad\vrule&
         \quad\hfil # \hfil\quad &
         \quad\hfil # \hfil\quad &
         \quad\hfil # \hfil\quad &
         \quad\hfil # \hfil\quad\vrule\cr
 & $0^+$ & $0^-$ & $1^+$ & $1^-$ \cr
 \noalign{\hrule}
 cb & 6800 & 6300   & 6730 & 6340   \cr
 ub & 5780 & 5277.6 & 5710 & 5331.3 \cr
 bs & 5890 & 5400   & 5820 & 5430\cr
 \noalign{\hrule} }}
\vfil
\break
\def\bsline#1#2#3#4#5#6{#1#2 & $#3\cdot 10^{#4}$ &
                        $\vert  V_{#5}V_{#6} \vert^2$ \cr}
\vbox{\halign{ # & # \cr
\bf Table 3.1 :& Branching ratios for two--body $\overline{B_s}$ decays
      \hfil \cr  \noalign{\vskip 1 mm}
     & assuming $a_1=1.11$, $a_2=0.21$,
      $\tau_{B_s}= 11.8\cdot 10^{-13}{\rm sec}$ , \cr
     & $V_{cb}=0.05$ and $V_{ub}=0.005$.\hfill \cr}}
\vskip 3 mm
\vbox{\offinterlineskip
\hrule
\halign{
\vrule\strut\quad # \quad\hfil\vrule &
            \quad # \quad\hfil\vrule &
            \quad # \quad\hfil\vrule\cr
\noalign{\hrule}
 \bsline{ \Dsxp}{ \Dsxm}{0.22}{-1}{cb}{cs}
 \bsline{ \Dsxp}{ \Aim} {0.16}{-1}{cb}{ud}
 \bsline{ \Dsp }{ \Rhom}{0.14}{-1}{cb}{ud}
 \bsline{ \Dsxp}{ \Rhom}{0.13}{-1}{cb}{ud}
 \bsline{ \Dsp }{ \Aim} {0.13}{-1}{cb}{ud}
 \bsline{ \Dsp }{ \Dsxm}{0.84}{-2}{cb}{cs}
 \bsline{ \Dsp }{ \Dsm} {0.76}{-2}{cb}{cs}
 \bsline{ \Dsp }{ \Pim} {0.55}{-2}{cb}{ud}
 \bsline{ \Dsxp}{ \Pim} {0.41}{-2}{cb}{ud}
 \bsline{ \Dsxp}{ \Dsm} {0.34}{-2}{cb}{cs}
 \bsline{ \Phii}{ \Jpsi}{0.13}{-2}{cb}{cs}
 \bsline{ \Dsxp}{ \Dmx} {0.11}{-2}{cb}{cd}
 \bsline{ \Dsp }{ \Kmx} {0.72}{-3}{cb}{us}
 \bsline{ \Dsxp}{ \Kmx} {0.71}{-3}{cb}{us}
 \bsline{ \Dsp }{ \Dmx} {0.46}{-3}{cb}{cd}
 \bsline{ \Dsp }{ \Km}  {0.42}{-3}{cb}{us}
 \bsline{ \Dsp }{ \Dm}  {0.40}{-3}{cb}{cd}
 \bsline{ \Dsxp}{ \Km}  {0.30}{-3}{cb}{us}
 \bsline{ \Kxz }{ \Dzx} {0.23}{-3}{cb}{ud}
 \bsline{ \Dsxp}{ \Dm}  {0.19}{-3}{cb}{cd}
 \bsline{ \Etap}{ \Jpsi}{0.15}{-3}{cb}{cs}
 \bsline{ \Etaa}{ \Jpsi}{0.13}{-3}{cb}{cs}
 \bsline{ \Kz  }{ \Dzx} {0.12}{-3}{cb}{ud}
 \bsline{ \Kz  }{ \Dz}  {0.83}{-4}{cb}{ud}
 \bsline{ \Kxp }{ \Dsxm}{0.67}{-4}{ub}{cs}
 \bsline{ \Etap}{ \Etac}{0.59}{-4}{cb}{cs}
 \bsline{ \Kxz }{ \Jpsi}{0.57}{-4}{cb}{cd}
 \bsline{ \Etaa}{ \Etac}{0.46}{-4}{cb}{cs}
 \bsline{ \Kxz }{ \Dz}  {0.45}{-4}{cb}{ud}
}
\hrule
}
\vfill\eject
\vbox{\offinterlineskip
\hrule
\halign{
\vrule\strut\quad # \quad\hfil\vrule &
            \quad # \quad\hfil\vrule &
            \quad # \quad\hfil\vrule\cr
\noalign{\hrule}
 \bsline{ \Kxp }{ \Aim} {0.43}{-4}{ub}{ud}
 \bsline{ \Phii}{ \Etac}{0.40}{-4}{cb}{cs}
 \bsline{ \Kp  }{ \Rhom}{0.39}{-4}{ub}{ud}
 \bsline{ \Kp  }{ \Aim} {0.37}{-4}{ub}{ud}
 \bsline{ \Kp  }{ \Dsxm}{0.34}{-4}{ub}{cs}
 \bsline{ \Kxp }{ \Rhom}{0.33}{-4}{ub}{ud}
 \bsline{ \Kp  }{ \Dsm} {0.23}{-4}{ub}{cs}
 \bsline{ \Phii}{ \Dzx} {0.14}{-4}{cb}{us}
 \bsline{ \Kz  }{ \Jpsi}{0.14}{-4}{cb}{cd}
 \bsline{ \Kp  }{ \Pim} {0.14}{-4}{ub}{ud}
 \bsline{ \Kxp }{ \Dsm} {0.12}{-4}{ub}{cs}
 \bsline{ \Kxp }{ \Pim} {0.10}{-4}{ub}{ud}
 \bsline{ \Kz  }{ \Etac}{0.49}{-5}{cb}{cd}
 \bsline{ \Etap}{ \Dzx} {0.39}{-5}{cb}{us}
 \bsline{ \Kxp }{ \Dmx} {0.33}{-5}{ub}{cd}
 \bsline{ \Etaa}{ \Dzx} {0.31}{-5}{cb}{us}
 \bsline{ \Phii}{ \Dz}  {0.30}{-5}{cb}{us}
 \bsline{ \Phii}{ \Dzxbar}{0.28}{-5}{ub}{cs}
 \bsline{ \Etap}{ \Dz}  {0.27}{-5}{cb}{us}
 \bsline{ \Etaa}{ \Dz}  {0.21}{-5}{cb}{us}
 \bsline{ \Kp } {\Kmx}  {0.20}{-5}{ub}{us}
 \bsline{ \Kxp }{ \Kmx} {0.18}{-5}{ub}{us}
 \bsline{ \Kp }{ \Dmx} {0.18}{-5}{ub}{cd}
 \bsline{ \Kxz }{ \Etac}{0.17}{-5}{cb}{cd}
 \bsline{ \Kp }{ \Dm}  {0.12}{-5}{ub}{cd}
 \bsline{ \Kp }{ \Km}  {0.11}{-5}{ub}{us}
 \bsline{ \Etap}{ \Dzxbar}{0.76}{-6}{ub}{cs}
 \bsline{ \Kxp }{ \Km}  {0.76}{-6}{ub}{us}
 \bsline{ \Kxz }{ \Aiz} {0.75}{-6}{ub}{ud}
 \bsline{ \Kz  }{ \Rhoz}{0.70}{-6}{ub}{ud}
 \bsline{ \Kz  }{ \Omeg}{0.70}{-6} {ub}{ud}
 \bsline{ \Kz  }{ \Aiz} {0.67}{-6}{ub}{ud}
 \bsline{ \Kxp }{ \Dm}  {0.65}{-6}{ub}{cd}
}
\hrule
}
\vfill\eject
\vbox{\offinterlineskip
\hrule
\halign{
\vrule\strut\quad # \quad\hfil\vrule &
            \quad # \quad\hfil\vrule &
            \quad # \quad\hfil\vrule\cr
\noalign{\hrule}
 \bsline{ \Etaa}{ \Dzxbar}{0.60}{-6}{ub}{cs}
 \bsline{ \Kxz }{ \Omeg}{0.60}{-6} {ub}{ud}
 \bsline{ \Kxz }{ \Rhoz}{0.59}{-6}{ub}{ud}
 \bsline{ \Phii}{ \Dzbar}{0.58}{-6}{ub}{cs}
 \bsline{ \Etap}{ \Dzbar}{0.53}{-6}{ub}{cs}
 \bsline{ \Etaa}{ \Dzbar}{0.40}{-6}{ub}{cs}
 \bsline{ \Kz  }{ \Etaa}{0.26}{-6} {ub}{ud}
 \bsline{ \Kz  }{ \Piz} {0.25}{-6} {ub}{ud}
 \bsline{ \Kxz }{ \Etaa}{0.18}{-6} {ub}{ud}
 \bsline{ \Kxz }{ \Piz} {0.18}{-6} {ub}{ud}
 \bsline{ \Kxz }{ \Dzxbar}{0.12}{-6} {ub}{cd}
 \bsline{ \Kz  }{ \Etap}{0.94}{-7} {ub}{ud}
 \bsline{ \Kz  }{ \Dzxbar}{0.63}{-7}{ub}{cd}
 \bsline{ \Kxz }{ \Etap}{0.62}{-7} {ub}{ud}
 \bsline{ \Phii}{ \Aiz} {0.49}{-7}{ub}{us}
 \bsline{ \Kz  }{ \Dzbar}{0.42}{-7}{ub}{cd}
 \bsline{ \Phii}{ \Omeg}{0.39}{-7}{ub}{us}
 \bsline{ \Phii}{ \Rhoz}{0.39}{-7}{ub}{us}
 \bsline{ \Etap}{ \Rhoz}{0.24}{-7}{ub}{us}
 \bsline{ \Etap}{ \Omeg}{0.24}{-7}{ub}{us}
 \bsline{ \Kxz }{ \Dzbar}{0.23}{-7}{ub}{cd}
 \bsline{ \Etap}{ \Aiz} {0.22}{-7}{ub}{us}
 \bsline{ \Etaa}{ \Rhoz}{0.18}{-7}{ub}{us}
 \bsline{ \Etaa}{ \Omeg}{0.18}{-7}{ub}{us}
 \bsline{ \Etaa}{ \Aiz} {0.17}{-7}{ub}{us}
 \bsline{ \Etaa}{ \Etaa}{0.13}{-7}{ub}{us}
 \bsline{ \Phii}{ \Etaa}{0.12}{-7}{ub}{us}
 \bsline{ \Phii}{ \Piz} {0.12}{-7}{ub}{us}
 \bsline{ \Etap}{ \Piz} {0.86}{-8}{ub}{us}
 \bsline{ \Etaa}{ \Piz} {0.65}{-8}{ub}{us}
 \bsline{ \Etap}{ \Etap}{0.63}{-8}{ub}{us}
 \bsline{ \Phii}{ \Etap}{0.41}{-8}{ub}{us}
 \bsline{ \Etaa}{ \Etap}{0.20}{-8}{ub}{us}
\noalign{\hrule}
} 
} 
\vfil\break
\vfil
\vbox{\halign{ # & # \cr
\bf Table 4.1 :&The contribution of several two--body channels to \cr
&the branching ratio $B_s\to$ stable charged particles only. \cr }}
\vskip 3 mm
\vbox{\offinterlineskip\hrule\halign{
  \vrule\strut\quad # \quad\hfil&\vrule \quad $ #\cdot 10^{-4}$
   \quad \vrule\cr
  \Dsp\Aim      & 4.37 \cr
  \Dsp\Pim      & 3.70 \cr
  \Phii\Jpsi    & 0.88 \cr
  \Dsp\Dsm      & 0.34\cr
  \Dsp\Km       & 0.28 \cr
  \Phii$\psi^\prime$&  0.24   \cr
  \Kp\Aim       & 0.19 \cr
  \Kp\Pim       & 0.14 \cr
  \Dsp\Kxm      & 0.11 \cr
\noalign{\hrule}
  Sum           &  10.3\cr
\noalign{\hrule}}}
\vfil\eject\end